\documentclass[12pt, a4paper]{article}
\usepackage{graphicx, caption, subcaption, braket, amsmath, amsfonts, tensor, comment}
\usepackage[dvipsnames]{xcolor}

\usepackage[normalem]{ulem}

\usepackage{jheppub}

\preprint{LITP-26-07\\ \hspace*{10.25cm} USTC-ICTS/PCFT-26-37}

\title{Low-Temperature Holographic Conductivity}

\author[1, 2]{Sabyasachi Maulik,}
\emailAdd{mauliks@ustc.edu.cn}

\author[3]{Leopoldo A. Pando Zayas,}
\emailAdd{lpandoz@umich.edu}

\author[3]{Jingchao Zhang}
\emailAdd{jingchaz@umich.edu}

\affiliation[1]{Interdisciplinary Center for Theoretical Study, University of Science and Technology of China, Hefei, Anhui 230026, China.}
\affiliation[2]{Peng Huanwu Center for Fundamental Theory, Hefei, Anhui 230026, China}
\affiliation[3]{Leinweber Institute for Theoretical Physics, University of Michigan, Ann Arbor, MI 48109, USA}

\abstract{We revisit holographic electrical conductivity in the regime of very low temperatures where quantum fluctuations in the throat of the near-extremal, asymptotically AdS$_4$ dual  black brane  are strongly coupled. We treat the fluctuations via an effective two-dimensional action capturing the effects of the Schwarzian modes at the scale $1/C$. Our main result is a non-monotonic temperature dependence: the quantum-corrected conductivity decreases as the temperature is lowered, reaches a minimum at $(CT_{\min}\simeq 0.023)$, and grows as $(1/\sqrt{CT})$ for $(CT\ll 1)$. We further estimate the one-loop contribution to the conductivity directly from the four-dimensional gravitational path integral and find qualitative agreement with the effective two-dimensional description in the regime $(CT\gg 1)$. We find that quantum corrections substantially modify the low-temperature behavior of the holographic conductivity in both approaches.}

\begin{document}

\maketitle
%%%%%%%%%%%%%%%%%%%%%%%%%%%%%%%%%%%%%%%%%%%%%%%%%%%%%%%%%%
\section{Introduction}

The AdS/CFT correspondence \cite{Maldacena:1997re, Gubser:1998bc, Witten:1998qj} has provided a pathway to address many phenomena in strongly coupled field theories \cite{Natsuume:2014sfa, Zaanen:2015oix, Baggioli:2019rrs}. Generically, the dual gravitational configuration requires a black hole at some finite temperature, $T$, and possibly other fields that depend on the physics that one aims to capture.  

It has recently been understood that low temperature near-extremal black holes behave, in many respects, like ordinary quantum systems. They are subject to strong quantum fluctuations at very low temperatures and to understand their behavior one needs to treat them in a paradigm where the temperature is considered small but non-vanishing, that is, one needs to respect the correct order of limits \cite{Iliesiu:2020qvm, Heydeman:2020hhw, Boruch:2022tno, Turiaci:2023wrh}. The description of these gravitational quantum fluctuations relies on the understanding of certain pseudo Goldstone modes and is, therefore, completely under control. Generic quantum gravitational fluctuations can be neglected given that curvatures are low in this regime. There has been a series of works discussing the implications of these quantum fluctuations for the thermodynamics of near-extremal black holes  \cite{Iliesiu:2022onk, Banerjee:2023quv, Kapec:2023ruw, Rakic:2023vhv, Banerjee:2023gll, Maulik:2024dwq, Harksen:2024uik, Kolanowski:2024zrq, Arnaudo:2024bbd,Arnaudo:2024rhv, Maulik:2025phe, Blacker:2025zca, Arnaudo:2025btb,PandoZayas:2026vbg,Alvarado:2026kio,Cheng:2026urz}, Hawking radiation \cite{Bai:2023hpd,Brown:2024ajk, Maulik:2025hax, Lin:2025wof, Bhattacharjee:2025wfv, Luo:2026epp} and absorption cross-sections \cite{Emparan:2025sao, Biggs:2025nzs, Emparan:2025qqf,Betzios:2025sct,Biggs:2026zlp}. Implications of these quantum fluctuations for aspects of the AdS/CFT correspondence have also been addressed \cite{Daguerre:2023cyx,Liu:2024gxr}. In particular,  an interesting novel behavior of shear viscosity to entropy density ratio, $\eta/s$,  has recently been reported \cite{Nian:2025oei,PandoZayas:2025snm,Cremonini:2025yqe,Gouteraux:2025exs,Kanargias:2025vul}.

{\it In this manuscript, we incorporate these quantum corrections in the context of  holographic electrical  conductivity.}  Previous relevant attempts to compute zero temperature transport, including electrical conductivity \cite{Edalati:2009bi,Edalati:2010hk, Edalati:2010pn}  as well as a recent insightful revision of the correlator at low temperatures \cite{Gouteraux:2025kta} took the temperature to zero first and worked directly with the exact extremal background containing an  AdS$_2$ throat. 

Operationally, there are two main approaches to including low-temperature quantum corrections in the context of the AdS/CFT correspondence. One approach is based on the effective low energy theory describing would-be massless modes (pseudo-Goldstone modes) in two dimensions; this is precisely the approach pursued in \cite{Liu:2024gxr,Cremonini:2025yqe,Gouteraux:2025exs,Kanargias:2025vul}. Another approach directly focuses on the four-dimensional gravitational path integral  and treats the would be zero modes in the higher dimensional theory; this is the framework  followed in \cite{PandoZayas:2025snm}. In this manuscript, we tackle the computation of holographic conductivity following both approaches. The four-dimensional approach, being a one-loop computation in the path integral, is purported to be the first correction to the semiclassical result. In the case of the partition function, this approach is sufficient to arrive at the same logarithmic-in-temperature correction to the entropy as with the effective Schwarzian approach \cite{Iliesiu:2022onk,Kapec:2023ruw,Rakic:2023vhv, Maulik:2024dwq,PandoZayas:2026vbg}. In the context of correlation functions, as is the case of interest here, one can only reasonably hope for the one-loop computation to uncover the first deviation from the semiclassical result.  Arguably, the two-dimensional approach, being based on properties of the Schwarzian theory, has a regime of validity that extends to lower temperatures. Therefore, in this manuscript we will rely mostly on the two-dimensional effective approach and will only briefly describe the results of the four-dimensional gravitational path integral approach. Our attitude is that both approaches have some regime of validity and here, rather than clarifying which one fits best in the holographic framework, we want to explore the implications of each approach for holographic transport at low temperature. In a sense, we explore the landscape of possible quantum corrections which we believe is an interesting question in itself.

The rest of the manuscript is organized as follows.  In Section \ref{Sec:Background} we review the standard computation for the holographic DC conductivity at zero temperature. Section \ref{sec:quantum_from_2D} contains our main result which is the revisited low-temperature holographic conductivity which includes the contribution of the strongly coupled Schwarzian modes in the throat region. Section \ref{Sec:4d-oneloop} contains a discussion of the four-dimensional approach to holographic conductivity via the  gravitational  path integral at one-loop. We conclude in Section \ref{Sec:Conclusions}. We relegate some technical details in the evaluation of the Wightman Green's function and zeta-function regularization to appendices \ref{App:Greens} and \ref{App: 4D details}, respectively.

%%%%%%%%%%%%%%%%%%%%%%%%%%%%%%%%%%%%%
\section{Holographic DC conductivity}\label{Sec:Background}

\subsection{Reissner-Nordstr\"om black brane}
    We consider the $\left(3+1\right)$-dimensional Einstein-Maxwell action with a negative cosmological constant $\Lambda = - \frac{3}{L^2}$
    \begin{equation}
    \label{eq:action}
        I = \frac{1}{2\kappa_{4}^{2}} \int d^{4}x \sqrt{-g} \left(R - 2\Lambda - L^2 F_{\mu\nu}F^{\mu\nu} \right).
    \end{equation}
    An electrically charged black brane solution is described by the following metric and gauge field
    \begin{subequations}
        \begin{align}
            ds^2 &= g_{\mu\nu} dx^{\mu} dx^{\nu} = \frac{r^2}{L^2} \left(-f(r) dt^2 + dx^2 + dy^2 \right) + \frac{L^2}{r^2 f(r)} dr^2, \label{eq:RN-AdS_solution} \\
            A &= \frac{Q}{L^2}\left(\frac{1}{r_{+}} - \frac{1}{r} \right) dt,\qquad f(r) = 1 - \frac{M}{r^3} + \frac{Q^2}{r^4},
        \end{align}
    \end{subequations}
    where $M$ and $Q$ are related to the mass and electric charge of the black brane. We consider the space spanned  by $\left(x, y\right)$ to be a two-torus with area $4\pi V_{2}$. The blackening factor $f(r)$ admits two real positive roots: $f\left(r_{\pm}\right) = 0$, $\left(r_{+} > r_{-}\right)$, which denote the outer and inner horizon radii of the black brane. The temperature and chemical potential, measured by an observer at asymptotic infinity, are given by the well-known expressions
    \begin{equation}
        T = \frac{4 r_{+}^{3} - M}{4 \pi L^{2} r_{+}^{2}},\quad \mu = \frac{Q}{L^{2} r_{+}}.
    \end{equation}

    Let us briefly discuss the near-extremal limit of the RN-AdS$_{4}$ solution. In the extremal limit, the locations of the two horizons coincide: $r_{+} = r_{-} = r_{0}$, and the temperature vanishes. For a very small Hawking temperature, $T$, it is possible to obtain the following Taylor expansions of the black hole parameters in powers of $T$
    \begin{subequations}
        \begin{align}
            r_{+} &= r_{0} +\frac{2\pi L^2}{3}\, T + \frac{2\pi ^2 L^4}{9 r_{0}}\, T^2 + \mathcal{O}\left(T^{3}\right),\\
            M &= 4 r_{0}^3 + 4 \pi L^2 r_{0}^{2}\, T + \frac{8 \pi^{2} L^{4} r_{0}}{3}\, T^{2} + \mathcal{O}\left(T^{3}\right),\\
            Q &= \sqrt{3} r_{0}^2 + \frac{2 \pi  L^2 r_{0}}{\sqrt{3}}\,T + \frac{2 \pi ^2 L^4}{3 \sqrt{3}}\, T^{2} + \mathcal{O}\left(T^{3}\right).
        \end{align}
    \end{subequations}
It is important that while performing the above small temperature expansion, we chose to keep the electric chemical potential, $\mu=\frac{\sqrt{3} r_{0}}{L^{2}}$ fixed at its extremal value. This is to be contrasted with the microcanonical discussion (at fixed electric charge), originally presented in \cite{Preskill:1991tb}, where the mass has no term linear in $T$, ultimately leading to a breakdown of the semiclassical thermodynamic approximation.\\
    
    The near-horizon geometry of the four-dimensional asymptotically AdS near-extremal Reissner-Nordstr\"{o}m black brane can be easily obtained by implementing the following coordinate transformation: 
    \begin{equation} \label{eq:near_horizon_scaling}
        r \to r_{+} + \frac{\pi L^2}{3} T \left(Y-1\right),\quad t \to \frac{t}{2\pi T}.
    \end{equation}
    In the limit $T \to 0$, the black brane solution in the new coordinates is given by a metric and gauge field (see also \cite{Banerjee:2023quv, Iliesiu:2022onk, Kanargias:2025vul}) of the form 
    \begin{align}
        ds^{2} = \left(g^{(0)}_{\mu\nu} + \lambda\, g^{(1)}_{\mu\nu}  \right) dx^{\mu} dx^{\nu},\quad A_{\mu} dx^{\mu} = \left(a^{(0)}_{\mu} + \lambda\,a^{(1)}_{\mu} \right) dx^{\mu},
    \end{align}
 where, for convenience, we have introduced a  parameter $\lambda$ proportional to the temperature:
    \begin{equation}\label{Eq:lambda}
      \lambda = \frac{2\pi L^2}{3 r_{0}} T.  
    \end{equation} The extremal limit  ($\lambda=0$) is given by the well-known AdS$_{2} \times \mathbb{T}^{2}$ geometry
	\begin{subequations}
		\begin{align}
			g_{\mu\nu}^{(0)} dx^{\mu} dx^{\nu} &= \frac{L^2}{6}\left(-\left(Y^2-1\right) dt^2 + \frac{dY^2}{Y^2-1} \right) + \frac{r_{0}^2}{L^2} \left(dx^2 + dy^2\right), \label{eq:RN-AdS4_NHE_metric}\\
			a_{\mu}^{(0)} dx^{\mu} &= \frac{Y - 1}{2\sqrt{3}} dt. \label{eq:RN-AdS4_NHE_gauge}
		\end{align}
	\end{subequations}
    Here $r_{0}$ is the event horizon radius of the extremal black brane. The radial coordinate $Y$ takes values in $Y \in \left[1, \infty\right)$, with $Y=1$ denoting the outer horizon of the full four-dimensional spacetime.
    The first order near-extremal corrections can be expressed as
    \begin{subequations}
		\begin{align}
			g_{\mu\nu}^{(1)} dx^{\mu} dx^{\nu} &= \frac{L^2}{9}\left(\left(Y + 1\right)^{2} \left(Y - 2\right) dt^2 + \frac{\left(Y - 2\right)\,dY^2}{\left(Y-1\right)^{2}} \right) + \frac{r_{0}^2 (Y+1)}{L^2} \left(dx^2 + dy^2\right), \label{eq:RN-AdS4_NHNE_metric}\\
			a_{\mu}^{(1)} dx^{\mu} &= -\frac{Y^{2} - 1}{4\sqrt{3}} dt. \label{eq:RN-AdS4_NHNE_gauge}
		\end{align}
	\end{subequations}

    We also observe that the semiclassical Bekenstein-Hawking entropy of the near-extremal black brane is given by
    \begin{equation}
        \frac{S_{\text{BH}}}{4\pi V_{2}} = \frac{r_{0}^{2}}{L^{2}} + \frac{2 r_{0}^{2}\,\lambda}{L^{2}} + \mathcal{O}\left(\lambda^{2}\right) = \frac{S_{\mathrm{ext}}}{4\pi V_{2}} + 4\pi^{2} \frac{T}{E_{\mathrm{gap}}} + \mathcal{O}\left(T^{2} \right).
    \end{equation}
    This linear in temperature correction, as explained in \cite{Preskill:1991tb}, implies that there exists a temperature scale, $E_{\mathrm{gap}} = \frac{3}{4 V_{2} r_{0}}$, below which the semiclassical thermodynamic description breaks down and quantum effects start to dominate \cite{Preskill:1991tb, Iliesiu:2020qvm, Iliesiu:2022onk, Banerjee:2023quv, Kapec:2023ruw, Rakic:2023vhv}.

    %%%%%%%%%%%%%%%%%%%%%%%%%
	%\subsection{Holographic DC conductivity from semiclassical Green's function}

    %%%%%%%%%%%%%%%%
    \subsection{Perturbation equations for the metric and gauge field}
    
	In the gauge/gravity duality, the boundary field theory charge current operator $J^{\mu}$ is dual to fluctuations of the bulk gauge field $A_{\mu}$ along the transverse directions, while the energy-momentum tensor $T^{\mu\nu}$ is dual to fluctuations of the metric $g_{\mu\nu}$ over the background \cite{Gubser:1998bc, Witten:1998qj}. To compute the conductivity, which relates a charge current to an applied electric field, we must consider coupled fluctuations of these bulk fields. A consistent set of perturbations involves turning on the following plane wave fluctuations in the bulk
	\begin{subequations}
    \begin{align}
        \delta g_{\mu\nu} dx^{\mu} dx^{\nu} &= \epsilon \int \frac{d\omega dk_{x} dk_{y}}{\left(2\pi\right)^3} e^{-i\omega t + i k_{x} x + i k_{y} y} h^{y}_{~t}\left(\omega, \vec{k}, r\right) g_{yy}\, dtdy \\
		\delta A_{\mu} dx^{\mu} &= \epsilon \int \frac{d\omega dk_{x} dk_{y}}{\left(2\pi\right)^3} e^{-i\omega t + i k_{x} x + i k_{y} y} a_{y}\left(\omega, \vec{k}, r\right) dy,
    \end{align}
	\end{subequations}
	where $\epsilon$ is an infinitesimal parameter that controls the strength of the perturbation. In this paper, we work in the hydrodynamic limit, in which we first take $\vec{k} \to 0$. One can then easily obtain the following linearized Maxwell's equation for $a_{y}\left(r\right)$ in the RN-AdS$_{4}$ geometry \eqref{eq:RN-AdS_solution}
    \begin{equation} \label{eq:ay_equation_coupled}
        a_{y}''(r) + \left(\frac{2}{r} + \frac{f'(r)}{f(r)} \right) a_{y}'(r) + \frac{\omega^{2} L^{4}}{r^{4} f(r)^{2}} a_{y}(r) + \frac{Q}{r^{2} L^{2} f(r)} h^{y\, \prime}_{~t}(r) = 0.
    \end{equation}
    The metric perturbation $h^{y}_{~t}\left(r\right)$ is determined by a constraint obtained from the $\left(r, y\right)$ component of linearized Einstein's field equations
    \begin{equation} \label{eq:hty_equation_full}
        \frac{dh^{y}_{~t}}{dr} + \frac{4L^2Q}{r^4} a_{y} = 0.
    \end{equation}
    Using the constraint Equation \eqref{eq:hty_equation_full} in Equation \eqref{eq:ay_equation_coupled}, we obtain the following decoupled ODE that governs the gauge field fluctuation
    \begin{equation} \label{eq:ay_equation_full}
        a_{y}''(r) + \frac{2r^4 + M r - 2Q^2}{r\left(Q^2 - M r + r^4\right)} a_{y}'(r) + \left(\frac{\omega^2 r^4 L^4}{\left(Q^2 - M r + r^4\right)^2} - \frac{4 Q^2}{r^2\left(Q^2 - M r + r^4\right)} \right) a_{y}(r) = 0.
    \end{equation}
    We solve these two equations by the method of matched asymptotics. We first divide the bulk spacetime into two regions:
    \begin{itemize}
        \item \textbf{Inner region:} This is the region very near to the horizon, where $\frac{r-r_{+}}{r_{+}} \ll 1$. The background solution in this region is very well approximated by equations \eqref{eq:RN-AdS4_NHE_metric} and \eqref{eq:RN-AdS4_NHE_gauge}.
        \item \textbf{Outer region:} In the outer region $\frac{\left(r-r_{+}\right)^2}{L^2 r_{+} \omega} \gg 1$. In this region, the effect of the black brane is negligible and the spacetime is well approximated by a four-dimensional AdS geometry.
    \end{itemize}
    Finally, we match the two asymptotic solutions in the overlapping region $\sqrt{L^{2} r_{+} \omega} \ll r - r_{+} \ll r_{+}$ to fix the undetermined constants and obtain the real-space two-point correlator $\langle J_{y} J_{y} \rangle$.

    %%%%%%%%%%%%%%%%\
    \subsection{Solution for photon perturbation}
    \subsubsection*{Solution in the inner region}

    In the near-horizon or inner region of the spacetime, we find it convenient to use the radial coordinate $Y$ of the AdS$_{2}$ geometry in \eqref{eq:RN-AdS4_NHE_metric}. Upon using the coordinate transformation
    \[r \to r_{+}\left(T\right) + \frac{\pi L^2}{3} T \left(Y-1\right),\quad t \to \frac{t}{2\pi T}\,, \]
    for very small $T$, and doing a Taylor expansion in powers of $T$, we find that at the leading order $O\left(T^{0}\right)$ the equation of motion for $a_{y}(Y)$ \eqref{eq:ay_equation_full} in the near-horizon region is given by
	\begin{equation} \label{eq:ay_eom_inner_I}
		\left(Y^2 - 1 \right) \frac{d^2 a_{y}}{dY^2} + 2 Y \frac{d a_{y}}{dY} + \left(-2 + \frac{\Omega^2}{1-Y^2}\right) a_{y} = 0,
	\end{equation}
	where $\Omega = \frac{\omega}{2\pi T}$. Since we are ultimately interested in using the Kubo formula, here and henceforth we consider only the leading terms in the limit $\Omega \to 0$. The last equation is an example of the associated Legendre differential equation. After imposing the ingoing boundary condition at the horizon $\left(Y \to 1\right)$, we find that an appropriate solution is given by
	\begin{equation} \label{eq:photon_soln_inner}
		a_{y, \text{in}}(Y) = \tilde{C}_{\text{in}}\, \left(\frac{Y-1}{Y+1}\right)^{-\frac{i \Omega}{2}} \left(Y - i \Omega\right).
	\end{equation}
	It is possible to calculate order ${\cal O}(T)$ corrections over the solution, but we do not consider them because their effect in the final result is subleading.

    % We will obtain the boundary data for the perturbation by matching this solution with the solution for the gauge field in the far from horizon region. To this end, it is convenient to rewrite \eqref{eq:photon_soln_inner} by rescaling the constant coefficient
    % \begin{equation}
    %     a_{y, \text{in}}(Y) = C_{\text{in}}\, T\, \left(\frac{Y-1}{Y+1}\right)^{-\frac{i \Omega}{2}} \left(Y - i \Omega\right).
    % \end{equation}
    Near the matching region, $Y \to \infty$, the perturbation behaves as
    \begin{equation} \label{eq:ayin_overlap}
        \left.a_{y, \text{in}}(Y)\right|_{Y \to \infty} = \tilde{C}_{\text{in}}\, \, Y +\frac{\tilde{C}_{\text{in}}  \Omega ^2}{2 Y} + \frac{i \tilde{C}_{\text{in}}  \left(\Omega ^2+1\right) \Omega }{3 Y^2}.
    \end{equation}

    %%%%%%%%%%%%%%%%
    \subsubsection*{Solution in the outer region}

    In the far from horizon region, we can ignore all ${\cal O}\left(\omega^2\right)$ terms in the linearized equation of motion. We follow the same conventions  as in \cite{Edalati:2009bi, PandoZayas:2025snm} and use $u = r/r_{0}$ as the radial coordinate in this region. Once again, using the small temperature expansion, we find that the leading order equation of motion takes the following form
    \begin{equation}
        a_{y}''(u) + \frac{2 \left(u^3+u^2+u+3\right)}{u \left(u^3+u^2+u-3\right)} a_{y}'(u) - \frac{12}{u^2 \left(u^4-4 u+3\right)} a_{y}(u) = 0,
    \end{equation}
    whose solution is 
    \begin{equation} \label{eq:photon_solution_outer}
    \begin{split}
        a_{y, \text{out}}\left(u\right) = a_{O} \left(1 - \frac{1}{u} \right) + b_{O} \Bigg(- \frac{41 + 5 u (13 u-20)}{108 u (u-1)^2} + \frac{17 (u-1) \tan ^{-1}\left(\frac{u+1}{\sqrt{2}}\right)}{324 \sqrt{2} u} \\ + \frac{7 \left(u - 1\right) \left( 2 \log \left(u-1\right) - \log \left(3 + u (2 + u)\right) \right)}{81 u} \Bigg).
    \end{split}
    \end{equation}
    We can obtain the asymptotic form of the outer solution $a_{y, \text{out}}\left(u\right)$ in the matching region $u \to 1$
    \begin{equation} \label{eq:ayout_overlap}
    \begin{split}
        \left. a_{y, \text{out}}(u)\right|_{u \to 1} = - &\frac{b_{O}}{18 (u-1)^2} - \frac{2 b_{O}}{9 (u-1)} - \frac{41 b_{O}}{108} + (u-1) \Big(a_{O} \\ & +\frac{1}{648} b_{O} \left(112 \log (u-1)+246-56 \log (6)+17 \sqrt{2} \tan ^{-1}\left(\sqrt{2}\right)\right) \Big).
    \end{split}
    \end{equation}
    Near the asymptotic boundary of AdS$_{4}$, \eqref{eq:photon_solution_outer} takes the form
    \begin{equation} \label{eq:ayout_boundary}
        \left. a_{y, \text{out}} \right|_{u \to \infty} = \left( a_{O} + \frac{17 \pi}{648 \sqrt{2}}b_{O} \right) - \left(a_{O} + \left(1 + \frac{17 \pi}{648 \sqrt{2}}\right) b_{O} \right)\frac{1}{u}.
    \end{equation}
    
    %%%%%%%%%%%%%%%%
    \subsection{Solution for metric perturbation}
    
    %%%%%%%%%%%%%%%%    
    \subsubsection*{Solution in the inner region}
    
    The metric perturbation $h^{y}_{~t} (r)$ is governed by the $\left(r, y\right)$ component of the linearized Einstein's equations, which is in fact the first order ODE \eqref{eq:hty_equation_full}
    \begin{equation}
        \frac{dh^{y}_{~t}}{dr} + \frac{4L^2Q}{r^4} a_{y} = 0.
    \end{equation}
    We can derive the near-horizon equation of motion for this field using the same coordinate transformation as in Equation \eqref{eq:near_horizon_scaling},  obtaining  
    \begin{equation}
        \frac{dh^{y}_{~t}}{dY} + \frac{4\pi L^4}{\sqrt{3}\,r_0^{2}}\, T \left(1 + O(T) \right) a_y(Y) = 0.
    \end{equation}
    In the inner region, we can consider $a_{y}(Y) = a_{y, \text{in}} (Y)$, where $a_{y, \text{in}} (Y)$ is the solution for the gauge field perturbation in the near-horizon region, as given in Equation \eqref{eq:photon_soln_inner}. Therefore, the solution for $h^{y}_{~t} (Y)$ in the inner region can be directly written as
    %%%%%%%%%%%%
    \iffalse
    \begin{equation} \label{eq:hty_solution_inner}
        h^{y}_{~t, \text{in}} \left(Y\right) = \frac{1}{2}C_{\mathrm{in}}T(Y^2-1)\;-\;\frac{i}{4}C_{\mathrm{in}}T(Y-1)\left(2+(Y+1)\log\left(\frac{Y-1}{Y+1}\right)\right)\Omega+H_{\text{in}},
    \end{equation}
    \fi
    %%%%%%%%%%%%
    \begin{equation}
    \label{eq:hty_solution_inner}
    \begin{aligned}
         h^{y}_{~t, \text{in}} (Y) =&   - \frac{4\pi L^{4}T}{\sqrt{3}\, r_{0}^{2}} \tilde{C}_{\text{in}} \int^{Y} dY' \left(Y' - i\Omega \right) \left(\frac{Y' - 1}{Y' + 1} \right)^{-\frac{i \Omega}{2}},\\
         =&-\frac{4\pi L^{4}T}{\sqrt{3}\,r_{0}^{2}}\tilde{C}_{\text{in}}
\Bigg[
\frac{(Y-1)(Y+1)}{2}
-\frac{i}{4}(Y-1)\Omega
\Big(
2+\log\frac{Y-1}{Y+1}\\
&+Y\log\frac{Y-1}{Y+1}
\Big)
+\mathcal{O}(\Omega^2)\Bigg]+H_{\rm in}.
    \end{aligned}
    \end{equation}
where $H_{\text{in}}$ is an integration constant.

    %%%%%%%%%%%%%%%%
    \subsubsection*{Solution in the outer region}
    Focusing on the region of the spacetime far from the outer horizon, and using the rescaled coordinate $u = \frac{r}{r_{0}}$, we obtain the equation
    \begin{equation}
        \frac{dh^{y}_{~t}}{du} + \frac{4\sqrt{3}\,L^2}{r_{0} u^4} a_{y} = 0.
    \end{equation}
    We can directly solve for $h^{y}_{~t}$ by using the solution \eqref{eq:photon_solution_outer} for $a_{y}(u)$ in the outer region; the result is
    \begin{equation}
    \begin{split}
        h^{y}_{~t, \text{out}}(u) = h_{O} &+ \frac{L^2}{648 \sqrt{3}\, r_{0} u^4} \Big(648\, a_{O} (4 u-3) - 17 \sqrt{2}\, b_{O} \left(u^4-4 u+3\right) \arctan \left(\frac{u+1}{\sqrt{2}}\right)\\ & \hspace{1 em} + 56\, b_{O} \left(-2 u^4 \log \left(1-u\right) + u^4 \log \left(3 + 2 u + u^2\right) \right. \\ & \hspace{2 em} \left. + 8 u \log \left(u-1\right) - 6 \log (u-1)+(3-4 u) \log (u (u+2)+3)\right)\\ & \hspace{3 em} + \frac{6 b_{O}}{u-1} \left(u \left(u \left(u \left(30-43 u \right) + 36 \right) - 218 \right) + 123 \right) \Big),
    \end{split}
    \end{equation}
    where $a_{O}$ and $b_{O}$ are the same constant coefficients as in Equation \eqref{eq:photon_solution_outer} and $h_{O}$ is a new integration constant. As with the other fluctuation, we find the way this solution behaves in the overlapping region $\left(u \to 1\right)$, and near asymptotic infinity $\left(u \to \infty\right)$
    \begin{align}
        \left. h^{y}_{~t, \text{out}} \right|_{u \to 1} &= h_{O} + \frac{L^2 (81 a_{O}+(45-14 i \pi ) b_{O})}{81 \sqrt{3} r_{0}}+\frac{5 b_{O} L^2 (u-1)}{9 \sqrt{3} r_{0}}-\frac{2 b_{O} L^2}{3 \sqrt{3} r_{0} (u-1)}. \label{eq:htyout_overlap} \\      
        %%%
        \begin{split}
        \left. h^{y}_{~t, \text{out}} \right|_{u \to \infty} &= h_{O} - \frac{\left(17 \sqrt{2}+224 i\right) \pi L^2  b_{O}}{1296 \sqrt{3}\, r_{0}} + \frac{L^2 \left(1296\, a_{O} + 17 \pi  \sqrt{2}\, b_{O}\right)}{324 \sqrt{3}\, r_{0} u^3}. \label{eq:htyout_boundary}
        \end{split}
    \end{align}
    
    %%%%%%%%%%%%%%%%
    \subsection{Matching in the overlapping region}

    In order to obtain the boundary data of the perturbation, we need to match the solutions obtained above in the `inner' and the `outer' regions. Towards this, we recall the relations \[r = r_{+} + \frac{\pi L^2}{3}T\left(Y-1\right),\quad u = \frac{r}{r_{0}},\] of the radial coordinates $Y$ and $u$ in the two regions; we can thus relate $Y$ and $u$ as
    \begin{equation} \label{eq:radial_coord_reln}
    Y = \frac{3 r_{0}}{\pi  L^2 T} (u-1) - \frac{\pi  L^2}{2 r_{0}} T.
    \end{equation}
    We use this relation to rewrite the asymptotic form in Equation \eqref{eq:ayin_overlap} in terms of $u$. The matching is performed in the overlap region defined by $Y \gg 1$, and $\lambda \ll (u-1) \ll 1$. It is convenient to introduce the dimensionless quantity
    \begin{equation}\label{Eq:Fraq-omega}
       \mathfrak{w} = \frac{2\pi L^{2}}{3 r_{0}}\,\omega.
    \end{equation}
    The matching relations depend crucially on the relative scaling of the parameters $\mathfrak{w}$ and $\lambda$. Although we are interested in the low-frequency $\left(\mathfrak{w} \to 0 \right)$, low-temperature $\left(\lambda \to 0 \right)$ behavior of the transport coefficients -- these two limits do not, in general, commute. This feature was already observed in \cite{Edalati:2009bi}, where the limit $\dfrac{\mathfrak{w}}{\lambda} \to \infty$ was considered in the computation of the conductivity; note that it corresponds to sending the temperature $T\to 0$ first. 
    
    In the following, we carefully consider both asymptotic regimes of the parameter space, namely $\dfrac{\mathfrak{w}}{\lambda} \to 0$ and $\dfrac{\mathfrak{w}}{\lambda} \to \infty$, and demonstrate how the final result depends on the choice of the limit. We then point out which regime is relevant for computing quantum corrections to the retarded correlator, which is our primary objective.

    %%%%%%%%%%%%%%%%
    \subsubsection*{Case I: $\dfrac{\mathfrak{w}}{\lambda} \ll 1, \lambda \to 0$}
    %%%%%%%%%%%%%%%%
    This case corresponds to sending $\omega\to 0$ first.   We start by rescaling $\tilde{C}_{\text{in}} \to \frac{\lambda}{2} C_{\text{in}}$, and rewriting Equation \eqref{eq:ayin_overlap} for the asymptotic form of $a_{y, \text{in}}$ in the overlapping region
    \begin{equation} \label{eq:ayin_overlap_iny_II}
        \left.a_{y, \text{in}}(Y)\right|_{Y \to \infty} = \frac{\lambda}{2} C_{\text{in}} Y + \frac{C_{\text{in}}  \mathfrak{w}^{2}}{16 \pi^{2} \lambda Y} + \frac{i C_{\text{in}} \mathfrak{w}  \left(\mathfrak{w}^{2} + 4\pi^{2}\lambda^{2} \right)}{48 \pi^{3} \lambda^{2} Y^2}.
    \end{equation}
    Next, we use the relation \eqref{eq:radial_coord_reln} to convert to $u$-coordinate, and perform a Taylor expansion in the small parameter $\dfrac{\mathfrak{w}}{\lambda}$, followed by an expansion in $\lambda$. Neglecting higher order terms, we arrive at
    \begin{equation} \label{eq:ayin_overlap_inu_new_I}
        \left.a_{y, \text{in}}(Y)\right|_{u \to 1} = C_{\text{in}} \left(\left(u - 1 \right) + \frac{i \mathfrak{w} \lambda^{2}}{48 \pi \left(u - 1\right)^2} \right) + \cdots.
    \end{equation}
    Comparing the two equations \eqref{eq:ayout_overlap} and \eqref{eq:ayin_overlap_inu_new_I}, we obtain the matching relations
 \begin{equation}
 \begin{aligned}
  \label{eq:matching_new_I}
    a_{O} &= C_{\text{in}} \left(1 + \frac{i \mathfrak{w} \lambda^{2}}{36 \pi^{3}} \left(\frac{41}{8} + \frac{17 \arctan \sqrt{2}}{24 \sqrt{2}} + \frac{7 \log \frac{u-1}{\sqrt{6}}}{3} \right) \right),\\
    b_O &= -\frac{3 i\, \mathfrak{w} \lambda^{2}}{8 \pi} C_{\text{in}}.
 \end{aligned}
 \end{equation}
 We emphasize that the matching relations obtained here do not reproduce the extremal result of \cite{Edalati:2009bi}. This is expected, because the above analysis corresponds to the regime $\mathfrak{w}/\lambda \ll 1$, i.e., $\omega \ll T$, compatible with sending $\omega\to 0$ first, whereas the extremal analysis effectively probes the opposite limit $\mathfrak{w}/\lambda \to \infty$ in which case $T\to 0$ first. The two orders of limits therefore access different regions of the parameter space, and need not commute.
 
 In particular, we have retained a term proportional to $\mathfrak{w}\lambda^{2}$ in the asymptotic expansion. Although this term vanishes in the strict extremal limit $\lambda \to 0$, it constitutes the leading contribution in the regime $\mathfrak{w} \ll \lambda$ and is therefore essential for a consistent matching in this limit. Dropping such terms would amount to taking the extremal limit prematurely and would obscure the non-commutativity of the $\mathfrak{w} \to 0$ and $\lambda \to 0$ limits.
    
    %%%%%%%%%%%%%%%%
    \subsubsection*{Case II: $\dfrac{\mathfrak{w}}{\lambda} \gg 1, \mathfrak{w} \to 0$}
    %%%%%%%%%%%%%%%%
    
    Let us now consider the opposite order of limits in the parameter space. Once again we start with Equation \eqref{eq:ayin_overlap_iny_II}, repeated here for convenience
    \begin{equation*}
        \left.a_{y, \text{in}}(Y)\right|_{Y \to \infty} = \frac{\lambda}{2} C_{\text{in}} Y + \frac{C_{\text{in}}  \mathfrak{w}^{2}}{16 \pi^{2} \lambda Y} + \frac{i C_{\text{in}} \mathfrak{w}  \left(\mathfrak{w}^{2} + 4\pi^{2}\lambda^{2} \right)}{48 \pi^{3} \lambda^{2} Y^2}.
    \end{equation*}
    As before, we convert everything to the $u$-coordinate using \eqref{eq:radial_coord_reln}, and perform an expansion in the regime $\dfrac{\lambda}{\mathfrak{w}} \ll 1$, with $\mathfrak{w} \to 0$ taken after the hierarchy is imposed. Neglecting higher order terms in $\dfrac{\lambda}{\mathfrak{w}}$ and $\mathfrak{w}$, we obtain the following asymptotic expansion of the inner solution in the overlapping region
    \begin{equation}
    \label{eq:ayin_overlap_inu_new_II}
         \left. a_{y, \text{in}} \right|_{u \to 1} = C_{\text{in}} \left( \left(u-1\right) + \frac{i \mathfrak{w}^{3}}{192 \pi^{3} \left(u-1\right)^{2}} \right) + \cdots.
    \end{equation}
 Comparing the two equations \eqref{eq:ayout_overlap} and \eqref{eq:ayin_overlap_inu_new_II}, we obtain the matching relations
 \begin{equation}
 \begin{aligned}
  \label{eq:matching_new_II}
    a_{O} &= C_{\text{in}} \left(1 + \frac{i \mathfrak{w}^{3}}{144 \pi^{3}} \left(\frac{41}{8} + \frac{17 \arctan \sqrt{2}}{24 \sqrt{2}} + \frac{7 \log \frac{u-1}{\sqrt{6}}}{3} \right) \right),\\
    b_O &= -\frac{3 i\, \mathfrak{w}^{3}}{32 \pi^{3}} C_{\text{in}}.
 \end{aligned}
 \end{equation}
This is exactly the matching relations derived in \cite{Edalati:2009bi} for an extremal RN-AdS$_{4}$ black brane, up to an overall normalization.\\
 
It was already noticed in \cite{Edalati:2009bi} that the photon perturbation $a_{y}$ behaves like a neutral scalar field of conformal dimension $\Delta = 2$ in the near-horizon region. This is evident from its equation of motion \eqref{eq:ay_eom_inner_I} and its asymptotic form near the AdS$_{2}$ conformal boundary given in equations \eqref{eq:ayin_overlap_inu_new_I} and \eqref{eq:ayin_overlap_inu_new_II}. According to the real-space AdS/CFT prescription of \cite{Son:2002sd, Policastro:2002se}, the retarded Green's function of the scalar operator $\mathcal{O}_{\Delta = 2}$ dual to this field in the IR CFT$_{1}$ is determined (up to an overall normalization) by the ratio of the coefficients of the normalizable to non-normalizable modes in the asymptotic expansion. Using the above results, we obtain
 \begin{equation}\label{Eq:GIR}
     \mathcal{G}_{\text{R, IR}} = 
        \begin{cases}
            i\mathfrak{w} \lambda^{2},\qquad &\mathfrak{w} \ll \lambda, \\
            i\mathfrak{w}^{3},\qquad &\mathfrak{w} \gg \lambda.
        \end{cases}
 \end{equation}
 On the other hand, the near-boundary behavior of the field in the full four-dimensional spacetime is dictated by Equation \eqref{eq:ayout_boundary}. The UV retarded Green's function, $G_{\text{UV},\,yy}^{\text{R}} \left(\omega, 0 \right)$ is given by the ratio of the normalizable and non-normalizable mode of the four-dimensional solution. Using our matching relations \eqref{eq:matching_new_I} and \eqref{eq:matching_new_II}, and the normalization of the Maxwell action for $a_{y}(u)$, we obtain in both cases\footnote{We ignore a numerical prefactor which is not relevant for the final result.}
 \begin{equation} \label{eq:retarded_correlator_uv_ir}
     \mathrm{Im}\, G_{\text{UV},\,yy}^{\text{R}} \left(\omega, 0 \right) = \frac{2 L^{2}}{\mu^{2} \kappa^{2}_{4}} \text{Im}\,\mathcal{G}_{\text{IR}}^{\text{R}} \left(\omega \right),
 \end{equation}
 with the appropriate IR Green's function This relationship will be used in later sections to determine the quantum-corrected Green's function from the two dimensional effective theory near the AdS$_{2}$ throat (also see \cite{Liu:2024gxr,Cremonini:2025yqe, Kanargias:2025vul}).

The holographic DC conductivity is obtained via the Kubo formula
\begin{equation}
    \mathrm{Re}\, \sigma_{\text{DC}} = \lim_{\omega \to 0} \frac{1}{\omega}\mathrm{Im}\, G_{\text{UV},\,yy}^{\text{R}} \left(\omega, 0 \right).
\end{equation}
We obtain
\begin{equation}\label{Eq:Cond-DC-2d}
    \mathrm{Re}\, \sigma_{\text{DC}} \sim
        \begin{cases}
            \frac{2L^2}{\kappa_{4}^{2}} \lim_{\omega \to 0} \left(\frac{T}{\mu} \right)^{2} = \frac{2L^2}{\kappa_{4}^{2}} \left(\frac{T}{\mu} \right)^{2},\quad &\omega \ll T.\\
            \frac{2L^2}{\kappa_{4}^{2}} \lim_{\omega \to 0} \left(\frac{\omega}{\mu} \right)^{2} = 0,\quad &\omega \gg T.
        \end{cases}
\end{equation}
The second line reproduces the familiar result for extremal charged black branes, originally obtained in \cite{Edalati:2009bi}. Note that this result is compatible with taking $T\to 0$ first. In contrast, in the regime $\omega \ll T$, $\sigma_{\text{DC}}$ scales as $T^{2}$, and therefore also vanishes if we subsequently impose  the extremal limit $T \to 0$. This computation demonstrates that, already at the classical level, the order of limits does not commute. \\
 
    We conclude this section with comments pertaining to the role of the metric fluctuations at this level in order to emphasize how it becomes more prominent in other quantum-corrected frameworks that we develop in subsequent sections. For the metric part, 
    % we use the condition that the source for metric perturbations is zero on the asymptotic boundary of the spacetime. Equating the $O(1)$ source term in equation \eqref{eq:htyout_boundary} to zero, we obtain
    % \begin{equation} \label{eq:hO_wrt_aObO}
    %     h_{O} = \frac{\pi L^2\left(17\sqrt{2} + 224 i \right)b_{O}}{1296\sqrt{3}\,r_{0}}.
    % \end{equation}
    % On the other hand, 
    matching the $u$-independent part in the overlapping region in Equation \eqref{eq:htyout_overlap} with the constant solution in Equation \eqref{eq:hty_solution_inner} we obtain
    \begin{equation} \label{eq:Hin_wrt_aObO_1}
       H_{\text{in}} = h_{O} + \frac{L^2 \left(81 a_{O} + \left(45 - 14 i \pi \right) b_{O} \right)}{81 \sqrt{3}\, r_{0}}.
    \end{equation}
    As a boundary condition at the horizon, we set $H_{in}=0$. Therefore, we obtain the relationship
    \begin{equation} \label{eq:Hin_wrt_aObO}
h_{O} =- \frac{L^2 \left(81 a_{O} + \left(45 - 14 i \pi \right) b_{O} \right)}{81 \sqrt{3}\, r_{0}}.
    \end{equation}    
With the above solution, we conclude that the metric perturbation decouples and it is completely determined by the perturbations of the photon, $a_y$.

\section{Low-temperature electrical conductivity from the Schwarzian theory} \label{sec:quantum_from_2D}

In this section, we approach low-temperature electrical conductivity by incorporating quantum corrections to the Green's function.  These corrections arise from the Schwarzian modes within the nearly AdS$_{2}$ throat, which appears close to the outer horizon at small temperatures. In this region, we may invoke the nearly-AdS$_{2}$/nearly-CFT$_{1}$ correspondence \cite{Almheiri:2014cka, Maldacena:2016upp,Jensen:2016pah}, and consider an operator $\mathcal{O}_{\Delta} (t)$ at the boundary of the AdS$_{2}$ geometry -- its bulk dual is the perturbation field $a_{y} (z)$ in the inner region. We subsequently couple this operator to the Schwarzian theory following the AdS/CFT correspondence.  We subsequently replace the effective two-dimensional throat region by the Schwarzian theory to which such an operator is coupled. Recall, from Equation \eqref{eq:ay_eom_inner_I}, that the linearized equation of motion of $a_{y}$ in this region can be expressed as
	\begin{equation} \label{eq:ay_eom_inner}
		\left(Y^{2} - 1\right) \frac{d^2a_{y}}{dY^2} + 2\, Y \frac{d a_{y}}{dY} + \left(-2 + \frac{\omega^2}{4\pi^2 T^2 \left(Y^{2} - 1\right)} \right) a_{y} = 0,
	\end{equation}    
which is identical to the equation of motion of a neutral, massive scalar field in two-dimensional anti-de Sitter geometry, written in the coordinate patch \eqref{eq:RN-AdS4_NHE_metric}. The usual holographic dictionary then determines that the conformal dimension of the dual operator $\mathcal{O}_{\Delta}$ is $\Delta = 2$.

We stress that although originally the gauge field perturbation $\delta A_{y}$ couples with the metric fluctuation $\delta g_{ty}$, the latter is completely determined by the constraint \eqref{eq:hty_equation_full} - which was used to arrive at the decoupled ODE \eqref{eq:ay_eom_inner}. In the inner region, the constraint Equation \eqref{eq:hty_equation_full} reduces to
\begin{equation}
    \frac{d h^{y}_{~t}(Y)}{dY} + \frac{4\pi L^{2} T}{\sqrt{3}\, r_{0}} a_{y}(Y) = 0.
\end{equation}

Coupling the operator ${\cal O}_\Delta(t)$ to the Schwarzian theory has the advantage of determining its two-point function. Our starting point is the exact frequency space Wightman function in the Schwarzian theory, which is given by \cite{Mertens:2022irh, Kanargias:2025vul}
	\begin{equation} \label{eq:Schwarzian_correlator_defn}
		\begin{split}
			\mathcal{G}^{\Delta} \left(\omega\right) &= \frac{1}{\pi} \int_{-\infty}^{\infty} dt\, e^{i\omega t} \langle \mathcal{O}_{\Delta} (t) \mathcal{O}_{\Delta} (0) \rangle, \\
			&= \frac{e^{S_{0}} \left(2 C\right)}{Z(T)\, 4\pi^{4} \left(2 C\right)^{2\Delta} \Gamma\left(2\Delta\right)} \int_{0}^{\infty} dk\, 2k \sinh \left(2\pi k\right) \sinh \left(2\pi \sqrt{k^2 + 2 C \omega}\right) e^{-\frac{k^2}{2 C T}},\\
			& \hspace{5 cm} \times \prod_{\sigma_{1, 2} = \pm 1} \Gamma \left(\Delta + i\sigma_{1} \sqrt{k^2 + 2 C \omega} + i \sigma_{2} k \right).
		\end{split}
	\end{equation}
	The two-point function is normalized by the exact partition function $Z(T)$ of the Schwarzian theory \cite{Mertens:2022irh}
	\begin{equation}
		Z(T) = \frac{\left(C T \right)^{\frac{3}{2}}}{\sqrt{2\pi}} \exp \left(S_{0} + 2\pi^2 C T \right),
	\end{equation}
where $C = \frac{1}{E_{\text{gap}}} = \frac{4 V_{2} r_{0}}{3}$ is the inverse coupling of the Schwarzian theory, and $S_{0}$ is the extremal semiclassical entropy.
	
We wish to evaluate the correlator for $\Delta = 2$, and study its low frequency behavior. However, for arbitrary parameter values, we are unable to evaluate the integral in \eqref{eq:Schwarzian_correlator_defn} exactly, and therefore turn to approximate methods, including numerical evaluation. One insightful approach by the authors of \cite{Kanargias:2025vul} found that depending on the relative strength of the three relevant scales $E_{\text{gap}}, \omega,$ and $T$ (or equivalently $1, C \omega, C T$), it is possible to split the analysis into different regions of the parameter space, and make some simplifying approximations to the functions appearing in \eqref{eq:Schwarzian_correlator_defn}. This way, it is possible to obtain an analytic estimate of the frequency space correlator in different parameter regimes. We follow the logic of \cite{Kanargias:2025vul}, and solve the integral \eqref{eq:Schwarzian_correlator_defn} in the region of parameter space where $C\omega \ll 1, C T, \left(C T\right)^2$; this is the quantum regime where the Schwarzian correction is expected to dominate the result.\\

\noindent Following \cite{Kanargias:2025vul}, let us rewrite the frequency space Wightman function as
	\begin{equation} \label{eq:Schwarzian_correlator_redefn}
		\begin{split}
			\mathcal{G}^{\Delta} \left(\omega\right) &= \frac{e^{S_{0}} \left(2 C\right)}{Z(T)\, 4\pi^{4} \left(2 C\right)^{2\Delta} \Gamma\left(2\Delta\right)} \int_{0}^{\infty} dk\, g_{T}(k) \times f\left(k, \omega\right),\\
			\text{where}\quad g_{T}(k) &= k\, e^{-\frac{k^2}{2 C T}} \sinh \left(2\pi k\right),\\
			f\left(k, \omega\right) &= \prod_{\sigma_{1, 2} = \pm 1} \Gamma \left(\Delta + i\sigma_{1} \sqrt{k^2 + 2 C \omega} + i \sigma_{2} k \right) \sinh \left(2\pi \sqrt{k^{2} + 2 C \omega} \right).
		\end{split}
	\end{equation}

\noindent The main observation of \cite{Kanargias:2025vul} was to note that integrals such as \eqref{eq:Schwarzian_correlator_redefn} can be well approximated in the quantum regime $C\omega \ll 1, C T, \left(C T\right)^2$ by replacing $f\left(k, \omega\right)$ with its large $k$ asymptote. Likewise, we use the large $k$ expansion shown in Equation \eqref{eq:f_large_k} and obtain an analytic estimate for the two-point Wightman function $\mathcal{G}^{\Delta = 2} \left(\omega \right)$. We relegate a number of details and checks regarding the evaluation of the Wightman function to Appendix \ref{App:Greens}. 
\iffalse
The integral can be approximated by
    \begin{equation} \label{eq:anest}
        \begin{split}
            \mathcal{G}^{\Delta = 2}\left(\omega\right) &= \frac{1}{24 C^{3}} \Bigg(C\omega \left(1 + 2 C\omega \right)^{2} + 4 C\omega \times C T \left(3 + 4 \pi^{2} C T \right) \\ & \hspace{5 em} + \left(\left(1 + 4 \pi^{2} C T \right) \left(2 C \omega  \left(2 \left(3 + \pi^{2} \right) C \omega + 9 \right) + 3 \right) \right. \\ & \hspace{6 em} \left. + 12 C T \left(8 \pi^{2} C T \left(2 \pi^{2} C T + 3 \right) + 3 \right) \right) \frac{\mathrm{erf} \left(\sqrt{2} \pi \sqrt{C T} \right)}{6 \pi^{2}} \\ & \hspace{7 em} + O\left(e^{-2\pi^{2} C T} \right) \Bigg).
        \end{split}
    \end{equation}
\fi
    The complete result is somewhat cumbersome, so we only report here the leading terms for low and high values of $CT$
	\begin{equation} \label{eq:anest_high_low_T}
        \mathcal{G}^{\Delta=2}\left(\omega\right) = \begin{cases}
		      \frac{\left(1 + 2 C \omega \right)^{2}}{6 C^{2}} \left( \frac{\omega}{4} + \frac{1}{\sqrt{2\pi^{3}C^{2}}} \left(1 + \frac{2C\omega\left(3 + 2\pi^{2} C \omega \right)}{3 \left(1 + 2 C \omega \right)^{2}} \right) \sqrt{CT} + \mathcal{O}\left(C T\right) \right), & C T \ll 1 \\
              \frac{4\pi^{2}}{3 C^{3}} \left(\left(C T\right)^{3} + \frac{3}{2\pi^{2}} \left(1 + \frac{\pi^{2} C \omega}{3} \right) \left(C T\right)^{2} + \mathcal{O}\left(C T\right) \right), & C T \gg 1
        \end{cases}
	\end{equation}
	
	\noindent The (imaginary part of the) retarded Green's function is obtained from the Wightman function by the formula
	\begin{equation} \label{eq:retarded_wightman_rel}
		\text{Im} \mathcal{G}_{\text{R, IR}} \left(\omega\right) = \frac{1}{2} \left(1 - e^{-\frac{\omega}{T}}\right) \mathcal{G}^{\Delta = 2} \left(\omega\right).
	\end{equation}
    In this way we obtain the retarded Green's function in the IR, {\it i.e,.},  in the effective two-dimensional theory near the AdS$_{2}$ throat. The retarded Green's function in the UV can then be obtained by using Equation \eqref{eq:retarded_correlator_uv_ir}.\\
    
\begin{figure}[t]
		\centering
		\begin{subfigure}{0.48\textwidth}
			\centering
			\includegraphics[width=\linewidth]{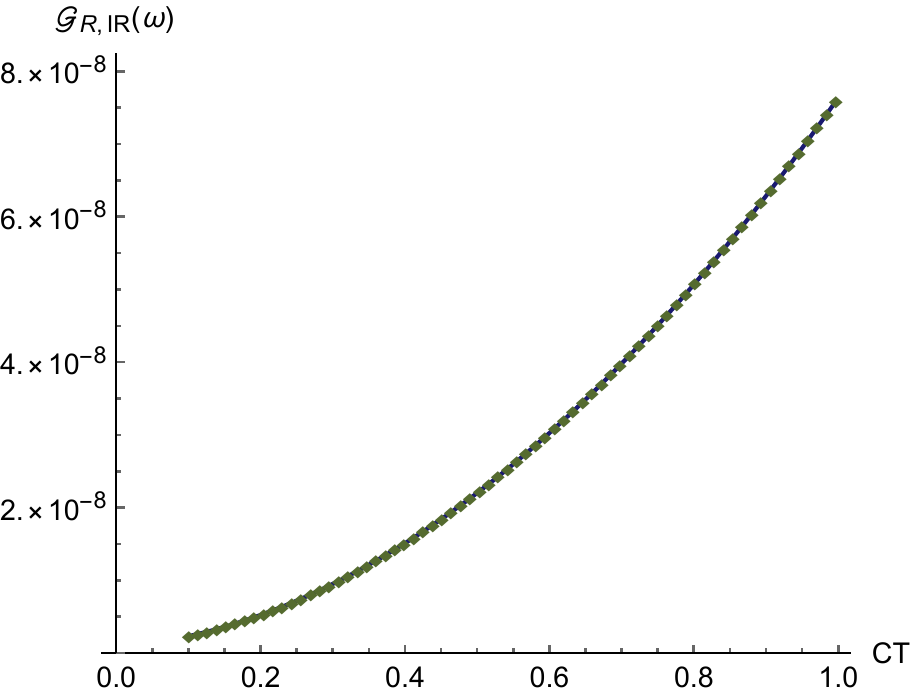}
			\caption{$C T \in \left[0.1, 1.0\right]$}
		\end{subfigure}
		\hfill
		\begin{subfigure}{0.48\textwidth}
			\centering
			\includegraphics[width=\linewidth]{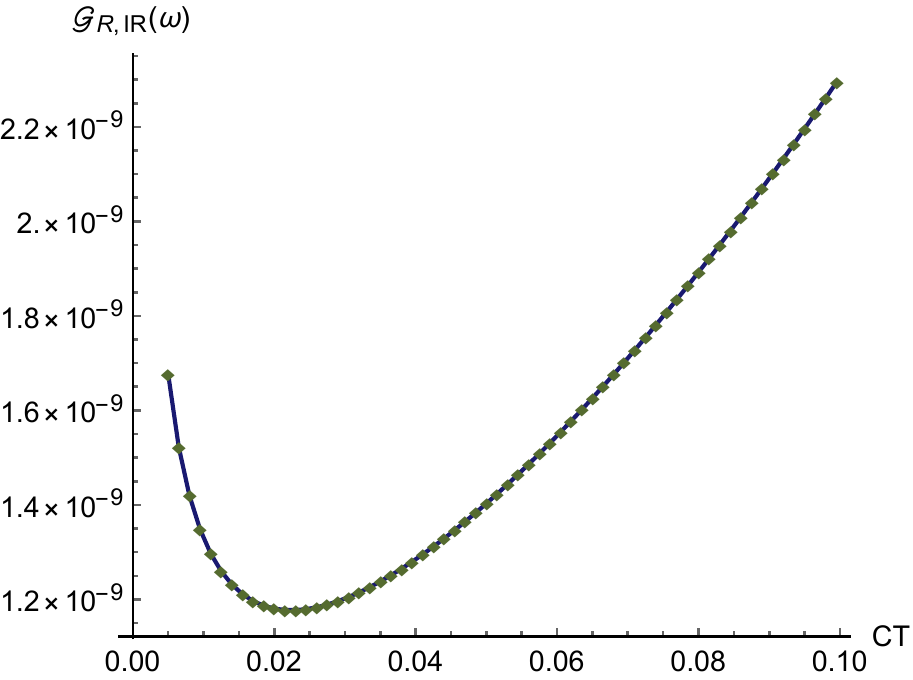}
			\caption{$C T \in \left[0.005, 0.1 \right]$}
		\end{subfigure}
		\caption{Comparison between numerical result and analytic estimate of the retarded Green's function Im $\mathcal{G}_{\text{R, IR}} \left(\omega\right)$. The solid line denotes the analytical curve, and the dots denote numerically obtained values. For numerical integration we choose $C\omega = 10^{-8}$.}
		\label{fig:RetardedG}.
	\end{figure}

	\noindent Using the approximate solution for the Wightman function $\mathcal{G}^{\Delta=2}\left(\omega\right)$ given in Equation \eqref{eq:full_schwarzian_correlator}, and Equation \eqref{eq:retarded_wightman_rel}, we obtain
	\begin{equation}\label{Eq:GIR-Schw}
		\text{Im} \mathcal{G}_{\text{R, IR}} \left(\omega\right) = 
        \begin{cases}
            \frac{\omega}{12 C^{2} \sqrt{2\pi^{3} C T}} + O\left(\omega^{2} \right), & C T \ll 1 \\
            \frac{\omega}{C^{2}} \left(\frac{2\pi^{2}}{3} \left(C T\right)^{2} + CT + \frac{3 + \pi^{2}}{24 \pi^{2}} + O\left(\frac{1}{C T} \right) \right) + O \left(\omega^{2} \right), & C T \gg 1
        \end{cases}
	\end{equation}
In Figure \ref{fig:RetardedG} we present  comparisons between analytical and numerical estimates for Im$\mathcal{G}_{\text{R, IR}}$. The observed error, $\left|G_{\text{R, IR}}^{(\text{num})} - G_{\text{R, IR}}^{(\text{est})} \right|$ is $O\left(10^{-15} \right)$ for the smallest $CT$ value considered in our numerical study, and it further decreases with increasing temperature. We provide more details of the precision of our approximation in Appendix \ref{App:Greens}. 
    %%%%%

\begin{figure}[t]
        \centering
        \includegraphics[width=0.6\linewidth]{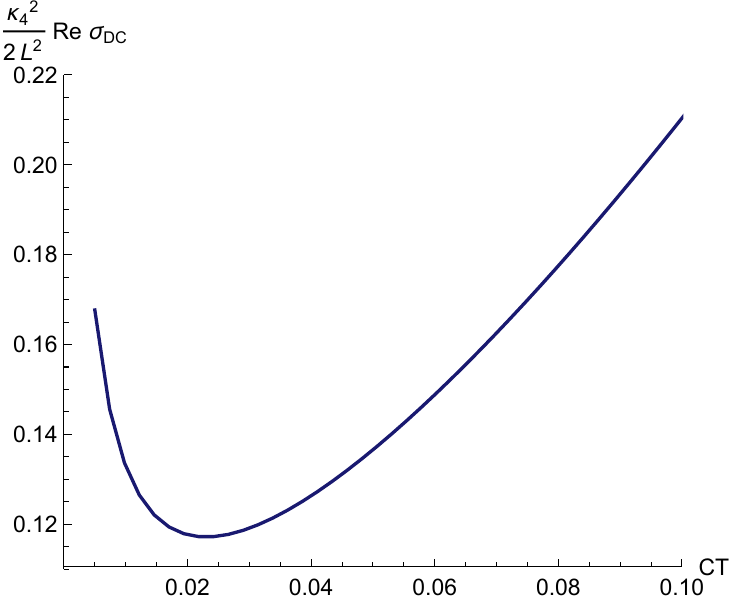}
        \caption{The quantum-corrected holographic conductivity at very low temperatures.}
        \label{fig:sigma_lowT}
    \end{figure}

\begin{figure}[t]
        \centering
        \begin{subfigure}{0.48\textwidth}
			\centering
			\includegraphics[width=\linewidth]{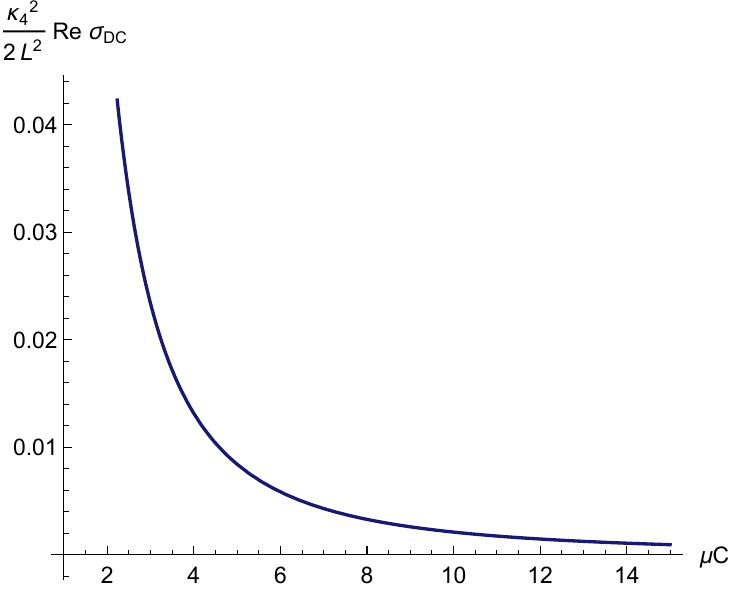}
			\caption{}
		\end{subfigure}
		\hfill
		\begin{subfigure}{0.48\textwidth}
			\centering
			\includegraphics[width=\linewidth]{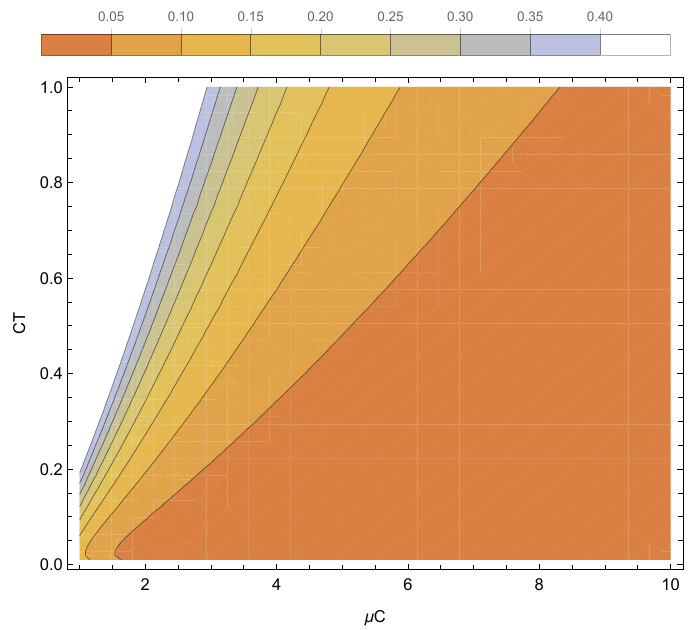}
			\caption{}
		\end{subfigure}
		\caption{Dependence of holographic DC conductivity $\left(\sigma_{\text{DC}} \right)$ on the chemical potential $\left(\mu \right)$: (a) For fixed temperature, the DC conductivity decreases monotonically and approaches zero for large $\mu$. (b) A contour plot of $\sigma_{\text{DC}}$ as a function of $CT$ and $\mu C$.}
		\label{fig:sigma_chempot_plt}.
    \end{figure}

To calculate the holographic conductivity, $\sigma_{\text{DC}}$, we first obtain the UV retarded Green's function using Equation \eqref{eq:retarded_correlator_uv_ir}, and use the Kubo formula
    \begin{equation}
        \mathrm{Re}\, \sigma_{\text{DC}} = \lim_{\omega \to 0} \frac{1}{\omega} \mathrm{Im}\, G^{\text{R}}_{\text{UV}, yy} \left(\omega, 0 \right).
    \end{equation}
    Ultimately, this leads to the following approximate formula for quantum-corrected DC conductivity at very low $\left(C T \ll 1 \right)$ temperatures
    \begin{equation}
        \mathrm{Re}\, \sigma_{\text{DC}} \simeq \frac{2 L^{2}}{\mu^{2}\kappa_{4}^{2}}\frac{1}{12 C^{2} \sqrt{2\pi^{3} C T}} \left(1 + \frac{2}{3} \left(\pi^{2} + 24 \right) C T - \frac{2}{15} \left(\pi^{2} - 240 \right) \left(C T\right)^{2} \right).
    \end{equation}
    We show the dependence of quantum-corrected holographic conductivity on temperature in Figure \ref{fig:sigma_lowT}. The conductivity derived above has a minimum at $T = T_{\text{min}}$, where $T_{\text{min}}$ is the unique positive solution of\footnote{Although $CT$ is small, the quadratic correction has a significant impact on the minimum, shifting the estimate from $CT_{\mathrm{min}} \simeq 0.044$ at linear order to $CT_{\mathrm{min}} \simeq 0.023$ when quadratic terms are included. We have further verified that including the cubic contribution changes the estimate of the minimum to $C T_{\mathrm{min}} \simeq 0.022$. This indicates that the perturbative expansion is converging and higher-order corrections are under control.}
    \begin{align}
        &6\pi^{2} \left(\pi^2 - 240 \right) \left(C T_{\text{min}} \right)^{2} - 10 \left(\pi^{2} + 24 \right) C T_{\text{min}} + 15 = 0,\nonumber \\
        \mathrm{or,}\quad &CT_{\rm min} = \frac{\sqrt{5 (2880 + 4560 \pi^2 - 13 \pi^4)}-120 - 5 \pi^2 }{6 \pi^2 (240 - \pi^2)},
    \end{align}
    \iffalse
    \begin{equation}
        C T_{\text{min}} = \frac{3}{2\left(\pi^{2} + 24 \right)},
    \end{equation}
    \fi
with the minimum value of the conductivity being given by
    \begin{equation}
        \text{Re}\, \sigma_{\rm DC}^{\rm min} = \frac{2L^2}{\kappa_{4}^{2}}\frac{1}{\mu^{2} C^2} \times 0.117194.
    \end{equation}
    %%%%%
    %%%%%
    The dependence of the DC conductivity on the chemical potential is illustrated in Figure \ref{fig:sigma_chempot_plt}. As shown in panel (a), for fixed temperature, the conductivity decreases with increasing $\mu$ and tends to zero in the large-$\mu$ regime. A more global picture is provided in panel (b), where we present a contour plot of $\sigma_{\mathrm{DC}}$
    in the $\left(\mu C, C T \right)$ plane. This representation makes the combined effect of temperature and chemical potential on the transport behavior manifest.
    %%%%%%
    %%%%
    
    In applications of the gauge/gravity duality, it is often of interest to study the optical conductivity, i.e., the frequency dependent conductivity. This quantity is defined analogously to the DC conductivity, but without any $\omega \to 0$ limit
    \begin{equation}
        \mathrm{Re}\, \sigma \left(\omega \right) = \frac{2 L^{2}}{\mu^{2} \kappa_{4}^{2}} 
        \frac{\text{Im}\, G^{\rm R}_{\text{UV}, yy}\left(\omega \right)}{\omega}.
    \end{equation}
    In Figure \ref{fig:optical_conductivity_frequency}, we present the behavior of the optical conductivity $\sigma\left( \omega\right)$ as a function of the frequency $\omega$, for several values of the temperature. For a given value of the temperature one always finds a minimum in the conductivity as a function of frequency.

    %%%%%%
    \begin{figure}[t]
        \centering
        \includegraphics[width=0.5\linewidth]{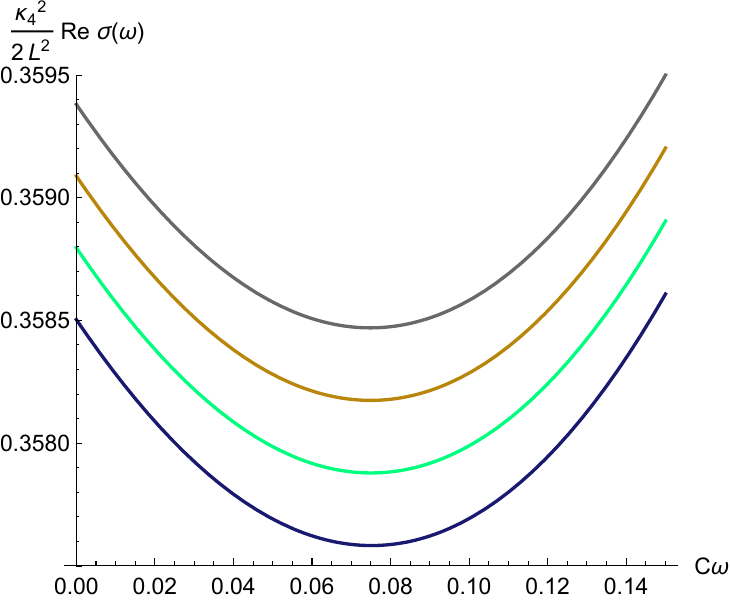}
        \caption{Frequency-dependent holographic optical conductivity, the temperature decreases from top to bottom.}
        \label{fig:optical_conductivity_frequency}
    \end{figure}

\section{One-loop holographic conductivity in higher dimensions}\label{Sec:4d-oneloop}

Strong quantum fluctuations localized in the throat of the near-extremal AdS$_4$ black brane have been shown to substantially affect thermodynamics. In the previous sections, we calculated the holographic conductivity using two-dimensional JT gravity, which is believed to control the dynamics of the near-horizon region of near-extremal black holes. In this section, we provide an alternative way to understand these corrections directly in higher-dimensional black holes by computing the gravitational path integral to one-loop order \cite{Iliesiu:2022onk, Kapec:2023ruw, Banerjee:2023quv, Rakic:2023vhv, Banerjee:2023gll, Maulik:2024dwq, Maulik:2025phe, Blacker:2025zca, PandoZayas:2026vbg}.

In this section, we explore holographic conductivity from the point of view of the gravitational path integral at one-loop order. Compared with the two-dimensional procedure used in previous sections, this higher-dimensional analysis is carried out only to one-loop order. It therefore captures the leading quantum effect as the temperature approaches zero while the Schwarzian effective theory is still weakly coupled. In other words, this higher-dimensional analysis has a narrower regime of validity than the two-dimensional approach, which uses the exact two-point function. The advantage of the higher-dimensional approach is that it is more direct, since no dimensional-reduction argument is needed. Comparing these two approaches is useful for understanding the relation between two-dimensional JT gravity and the full dynamics of a higher-dimensional near-extremal black hole, especially when an explicit dimensional reduction is not known. A four-dimensional analysis of holographic shear viscosity was previously reported in \cite{PandoZayas:2025snm}.

A comprehensive discussion of the one-loop Euclidean path integral around a near-extremal black hole and its zero modes can be found in \cite{Blacker:2025zca}. The use of this one-loop Euclidean path integral to calculate holographic correlators was established in \cite{PandoZayas:2025snm}.

%%%%%%%%%%%%%%%%%%%%%%
\subsection{Review of holography at one-loop level}
For the reader's convenience, we first briefly review the procedure. More details can be found in \cite{Blacker:2025zca, PandoZayas:2025snm}.

We start from the Euclidean path integral
\begin{equation}
    Z = \int [Dg]\,[DA] e^{-I[g,A]},
\end{equation}
where the action is given in \eqref{eq:action}, together with the appropriate boundary and gauge-fixing terms. Up to one-loop order, this partition function can be written as
\begin{align}
Z \approx& \exp\!\left(-I[\bar{g},\bar{A}]\right)
\int [Dh][Da]\,
\exp\!\Biggl[
-\int d^4x\,\sqrt{\bar{g}} \Bigl(
h^*\Delta_L[\bar{g},\bar{A}]h
+ a^*P[\bar{g},\bar{A}]a \nonumber \\
&\hspace{4.5cm}
+ \left(h^*O_{\mathrm{int}}[\bar{g},\bar{A}]a+\mathrm{h.c.}\right)
\Bigr)
\Biggr],
\end{align}
where $h$ and $a$ schematically denote perturbations of the metric and gauge field around the background. The quadratic operators can be found in Section 3.1 of \cite{Blacker:2025zca}. Compared with the saddle contribution, the one-loop part of the partition function is suppressed by Newton's constant and is therefore subleading. However, there are three types of ``would-be'' zero modes whose eigenvalues go to zero as the background configuration approaches extremality. Therefore, when the temperature is sufficiently small, the contribution of these zero modes becomes important. In this one-loop calculation, we neglect the nonzero modes. The zero modes can be found in Section 3.2 of \cite{Blacker:2025zca}; following the same terminology, we refer to them as tensor, vector, and gauge zero modes.

To calculate the holographic correlators, we take the background configuration, i.e., the saddle, to be composed of three parts:
\begin{equation}
\bar{g}=g^{(0)}+g^{(1)}+\delta g,\qquad \bar{A}=A^{(0)}+A^{(1)}+\delta A,
\end{equation}
where $\{g^{(0)},A^{(0)}\}$ is the near-horizon extremal configuration, $\{g^{(1)},A^{(1)}\}$ are the linear perturbations due to the small temperature, and $\{\delta g,\delta A\}$ are the holographic perturbations; for example, \eqref{eq:photon_soln_inner} and \eqref{eq:hty_solution_inner} are used to calculate the conductivity in this paper. The eigenvalues of the zero modes are lifted by both $\{g^{(1)},A^{(1)}\}$ and $\{\delta g,\delta A\}$; the latter reflects the coupling between the holographic perturbations and the zero modes.

Integrating the zero modes in the Gaussian approximation leads to the one-loop partition function, which is given by the product of the corresponding determinants:
\begin{equation}
\label{eq:z1 products}
    Z_1=Z_{\rm Tensor}\, Z_{\rm Gauge}\, Z_{\rm Vector_x}\,Z_{\rm Vector_y},
\end{equation}
where there are two sets of vector zero modes due to the two Killing vectors on $\mathbb{T}^2$. Using the prescription in \cite{PandoZayas:2025snm}, the one-loop contribution to the Green's functions is obtained from
\begin{equation}
    G^{\rm one-loop}(\omega)=\frac{\delta Z_1}{\delta\phi(\omega)\delta\phi(-\omega)}/Z_1,
\end{equation}
where $\phi$ is the source at the AdS$_4$ boundary. In the following, we show the calculation of the contribution from the tensor zero modes, i.e., $Z_{\rm Tensor}$. The contributions from the other zero modes can be obtained similarly, so we only show the results.

%%%%%%%%%%%%%%%
\subsection{Contributions from lifted zero modes}
In the near-horizon throat, there is a set of tensor zero modes whose eigenvalues under the Lichnerowicz operator vanish as the background temperature goes to zero. The explicit form of these tensor zero modes for the AdS$_4$ black brane is
\begin{equation}
h_{\mu\nu}^{(n)}\,dx^\mu dx^\nu
=
C_n e^{in\tau}
\left(
\frac{-1+Y}{1+Y}
\right)^{\frac{n}{2}}
\left(
-d\tau^2
+
2i\,\frac{d\tau\, dY}{Y^2-1}
+
\frac{dY^2}{(Y^2-1)^2}
\right),
\end{equation}
where $n\ge2$ and the $C_n$ are normalization constants. More details and general properties of the tensor zero modes can be found in \cite{Maulik:2024dwq, PandoZayas:2026vbg}. These tensor zero modes are the higher-dimensional avatars of the boundary reparameterization modes in two-dimensional JT gravity. Their contribution to the partition function becomes large at low temperatures. The coupling between these tensor zero modes and the holographic perturbations $\{\delta A_\mu,\delta g_{\mu\nu}\}$ leads to the quantum correction to the holographic conductivity. This coupling is captured by the lifted eigenvalues of the tensor zero modes.

Let us consider the eigenvalue equation of the Lichnerowicz operator. Let $h_{\alpha\beta}^{(n)}$ be the tensor zero modes in the extremal near-horizon solution. Because of the fluctuations mentioned above, the eigenvalues become nonzero; that is, they are lifted by acquiring small nonzero eigenvalues. The leading nonzero correction to the eigenvalue occurs at $O\left(\epsilon^2\right)$:
\begin{align}
    &\Delta_{L}^{\alpha\beta, \mu\nu}\left(g^{(0)}+Tg^{(1)}+\epsilon\,\delta g, A^{(0)} + TA^{(1)}+\epsilon\,\delta A \right)\,h^{(n)}_{\mu\nu} \nonumber \\
    &= \underbrace{\Delta_{L}^{\alpha\beta, \mu\nu}\left(g^{(0)}, A^{(0)} \right)\,h^{(n)}_{\mu\nu}}_{=0}+T\, \delta\Delta_{L}^{\alpha\beta, \mu\nu}(g^{(1)},A^{(1)}) \,h^{(n)}_{\mu\nu} + \epsilon^{2} \delta\Delta_{L}^{\alpha\beta, \mu\nu}(\delta g,\delta A)\,h^{(n)}_{\mu\nu}.
\end{align}
The second term comes from the small-temperature metric perturbations, $\{g^{(1)},A^{(1)}\}$, which lead to a lifted eigenvalue of
\begin{equation}
    \delta\Lambda^{\rm Tensor}_{n}(g^{(1)},A^{(1)}) =
\frac{2\pi l^{2}nT}{r_{0}}.
\end{equation}
This term is present even in the absence of the holographic perturbations relevant for calculating the conductivity in this paper. This lifted eigenvalue leads to the universal entropy correction of the form $\tfrac{3}{2}\log{T}$. We want to calculate the change in the eigenvalue of the Lichnerowicz operator due to the fluctuations $\{\delta A,\delta g\}$, namely,
\begin{equation}
    \label{eq: lifted eigenvalues, ag part}
\delta\Lambda^{\rm Tensor}_{n}(\delta g,\delta A) = \int d^{4}x \sqrt{-g^{(0)}} h_{\alpha\beta}^{(n)*}\,\delta\Delta_{L}^{\alpha\beta, \mu\nu}(\delta A,\delta g)\,h^{(n)}_{\mu\nu},
\end{equation}
where $\delta\Delta_{L}^{\alpha\beta, \mu\nu}$ is the change in the Lichnerowicz operator at leading order in $\epsilon$.

At leading order in temperature, the on-shell perturbations of the metric are given by
\begin{equation}
\label{eq: perturbed h}
    \begin{aligned}
    \delta g_{\mu\nu} dx^{\mu} dx^{\nu} &= \epsilon \int \frac{d\omega dk_{x} dk_{y}}{\left(2\pi\right)^3} e^{-i\omega t + i \vec{k}\cdot\vec{x}}\, h^{y}_{~t}\left(\omega, \vec{k}, Y \right) g_{yy}\, dt\,dy,\\
    &= \epsilon \int \frac{d\omega dk_{x} dk_{y}}{\left(2\pi\right)^3} e^{-\Omega \tau + i \vec{k}\cdot\vec{x}}\, h^{y}_{~t,\mathrm{in}}\left(\omega, \vec{k}, Y \right) g_{yy}\,\frac{dt}{d\tau} d\tau dy,
\end{aligned}
\end{equation}
where $h^{y}_{t,\mathrm{in}}(\omega,\vec{k},Y)$ can be replaced by the expression in \eqref{eq:hty_solution_inner} in the $\vec{k}\to0$ limit. In the following, we replace $\tilde{C}_{\text{in}}$ by $\tilde{C}_{\text{in}}(\omega,\vec{k})$ to emphasize its role as a Fourier coefficient. Similarly, the perturbations of the gauge field are given by
\begin{equation}
\label{eq: perturbed a}
\begin{aligned}
    \delta A_{\mu} dx^{\mu} &= \epsilon \int \frac{d\omega dk_{x} dk_{y}}{\left(2\pi\right)^3} e^{-i\omega t + i\vec{k}\cdot\vec{x}}\, a_{y}\left(\omega, \vec{k}, Y \right)\,dy,\\
    &= \epsilon \int \frac{d\omega dk_{x} dk_{y}}{\left(2\pi\right)^3} e^{-\Omega \tau + i\vec{k}\cdot\vec{x}}\, \tilde{C}_{\text{in}}\left(\omega, \vec{k}\right) \left(Y - i \Omega\right) \left(\frac{Y-1}{Y+1}\right)^{-\frac{i \Omega}{2}} dy,
\end{aligned}
\end{equation}
where the coefficient $\tilde{C}_{\text{in}}\left(\omega,\vec{k}\right)$ is related to the boundary source of the perturbation $\delta A_{\mu}$ by the relations in \eqref{eq:matching_new_I}. The calculation of the lifted eigenvalues $\delta \Lambda^{\rm Tensor}_n$ is similar to that performed in \cite{PandoZayas:2025snm}.

The resulting imaginary part of the lifted eigenvalues, which is relevant for the real part of the conductivity, is given by
\begin{equation} \label{eq:delta_lambda_final}
    \delta \Lambda^{\rm Tensor}_{n}(\delta g,\delta a) = -\epsilon ^2 \frac{ l^{2} L^{6} (1+n)\, T^{2}}
{3\pi^{2}\, r_{0}^{4}\, V}\int d\omega\, dk_{x}\, dk_{y}\;\omega \; C_{\rm in}\left(\omega, \vec{k} \right)\, C_{\rm in}\left(-\omega, -\vec{k}\right).
\end{equation}
Here $V$ is the volume of $\mathbb{T}^2$. We have taken the limit $\Omega\to0$. Up to terms of higher order in $\Omega$ and $T$, see \eqref{eq:matching_new_I}, the rescaled Fourier coefficient $C_{\rm in}(\omega,\vec{k})$ is the same as the boundary source $a_{O}(\omega,\vec{k})$. The variable $l$, which has dimensions of length, is introduced in the normalization of the tensor zero modes:
\begin{equation}
    1=\frac{1}{2\kappa^2 l^2}\int d^4x\sqrt{-g}\,h^{(n)*}_{\mu\nu}h^{(n)\mu\nu}.
\end{equation}

Similarly, we can calculate the lifted eigenvalues of the vector and gauge zero modes:
\begin{equation}
\delta\Lambda^{\rm Vector,x}_{n}(\delta g,\delta a) = \int d^{4}x \sqrt{-g^{(0)}} \hat{h}_{\alpha\beta}^{(n)*}\,\delta\Delta_{L}^{\alpha\beta, \mu\nu}(\delta A,\delta g)\,\hat{h}^{(n)}_{\mu\nu},\quad c_x=1, c_y=0,
\end{equation}
\begin{equation}
\delta\Lambda^{\rm Vector,y}_{n}(\delta g,\delta a) = \int d^{4}x \sqrt{-g^{(0)}} \hat{h}_{\alpha\beta}^{(n)*}\,\delta\Delta_{L}^{\alpha\beta, \mu\nu}(\delta A,\delta g)\,\hat{h}^{(n)}_{\mu\nu}, \quad c_x=0, c_y=1,
\end{equation}
\begin{equation}
\delta\Lambda^{\rm Gauge}_{n}(\delta g,\delta a) = \int d^{4}x \sqrt{-g^{(0)}} a_{\alpha}^{(n)*}\,\delta P^{\alpha\beta}(\delta A,\delta g)\,a^{(n)}_{\beta}.
\end{equation}
The resulting imaginary parts of these lifted eigenvalues are
\begin{equation}
\label{eq:lifted eigenvalue vectorx}
    \delta \Lambda^{\rm Vector,x}_{n}(\delta g,\delta a) =\epsilon^2
\frac{l^{2} L^{6} T^{2}}
{24\pi^2 r_{0}^{4}V}
\left[
4+n\Bigl(
3+8\gamma_{\mathrm{E}}
+8\ln 2
+8\ln\tilde{\epsilon}
+8\psi(n)
\Bigr)
\right]\mathrm{Int},
\end{equation}
\begin{equation}
\label{eq:lifted eigenvalue vectory}
    \delta \Lambda^{\rm Vector,y}_{n}(\delta g,\delta a) =\epsilon^2
\frac{l^{2}L^{6}T^{2}}
{24\pi^2 r_{0}^{4}V}
\left[
12+n\Bigl(
5+16\gamma_{\mathrm{E}}
+16\ln 2
+16\ln\tilde{\epsilon}
+16\psi(n)
\Bigr)
\right]\mathrm{Int},
\end{equation}
\begin{equation}
\label{eq:lifted eigenvalue gauge}
    \delta \Lambda^{\rm Gauge}_{n}(\delta g,\delta a) = -\epsilon ^2
\frac{l^{2} L^{6} n T^{2}}
{12\pi^{2} r_{0}^{4} V}
\left(
1+3\gamma_{\mathrm{E}}
+\ln 8
+3\ln \tilde{\epsilon}
+3\psi(n)
\right)\mathrm{Int},
\end{equation}
together with
\begin{equation}
 \delta \Lambda^{\rm Vector,x}_{n}(g^{(1)}, A^{(1)}) =\delta \Lambda^{\rm Vector,y}_{n}(g^{(1)}, A^{(1)})=
\frac{4\pi l^{2} n}{3 r_{0}},
\end{equation}
\begin{equation}
    \delta \Lambda^{\rm Gauge}_{n}(g^{(1)}, A^{(1)})=
\frac{2\pi l^{2} n}{3 r_{0}},
\end{equation}
where \(\psi(n)\equiv \Gamma'(n)/\Gamma(n)\) denotes the digamma function, and $\mathrm{Int}$ stands for the common integral term
\begin{equation}
    \int d\omega\, dk_{x}\, dk_{y}\;\omega \; C_{\rm in}\left(\omega, \vec{k} \right)\, C_{\rm in}\left(-\omega, -\vec{k}\right).
\end{equation}
Unlike the calculation of the tensor zero modes, the integral in the $Y$ direction in \eqref{eq:lifted eigenvalue vectorx}--\eqref{eq:lifted eigenvalue gauge} diverges as $Y\to\infty$. We introduce a boundary cutoff $Y\in[1,1/\tilde{\epsilon}]$ to regulate the integral. In extracting the physical conductivity, we drop the divergent $\ln \tilde{\epsilon}$ terms, assuming a minimal-subtraction scheme. The mixing operator $O_{\rm int}$ continues to have zero eigenvalues even after we turn on $\{g^{(1)},A^{(1)}\}$ and $\{\delta g,\delta A\}$, and it does not contribute to the final result for the conductivity.

%%%%%%%%%%%%%%%%%%%%%%%%%%%%%%%%%%%%%%%%%%%%%%
\subsection{One-loop conductivity}

After obtaining the lifted eigenvalues, we apply $\zeta$-function regularization to calculate the one-loop determinant:
\begin{equation}
\label{eq:infinite product}
    Z_{\rm one-loop}=\prod_{n\ge n_0}\left(\Lambda_n(g^{(1)},A^{(1)})+\Lambda_n(\delta g,\delta A)\right)^{-1},
\end{equation}
where $n_0=2$ for the tensor zero modes and $n_0=1$ for the vector and gauge zero modes. We apply the above formula separately to all three types of zero modes. The overall one-loop partition function is the product shown in \eqref{eq:z1 products}. We leave the details of the $\zeta$-function regularization to Appendix~\ref{App: 4D details}.

Following the standard AdS/CFT prescription, at tree level, the retarded Green's function should be given by the variation of the on-shell action:
\begin{equation}
    G_R^{\rm tree}(\omega)\sim\frac{\delta^2 Z_{\rm tree}}{\delta C_{\rm in}(\omega)\delta C_{\rm in}(-\omega)}/Z_{\rm tree}.
\end{equation}
The one-loop correction to the retarded Green's function is given by the variation of the one-loop determinant:
\begin{equation}
    G_R^{\rm one-loop}(\omega)\sim\frac{\delta^2 Z_{1}}{\delta C_{\rm in}(\omega)\delta C_{\rm in}(-\omega)}/Z_1,
\end{equation}
which gives the imaginary part of the retarded Green's function:
\begin{equation}
\label{eq: Im G}
    \operatorname{Im} G_R^{\rm one-loop}(\omega)=
\frac{\,L^{6}T\,\omega}
{48\pi^{3}r_{0}^{3}V}
\left(
-17
-7\gamma_{\mathrm{E}}
+22\ln 2
-15\ln 3
+7\ln\!\left(\frac{\pi l^{2}T}{r_{0}}\right)
+9\ln\tilde{\epsilon}
\right).
\end{equation}
The Kubo formula then gives
\begin{equation}
    \operatorname{Re}\,\sigma_{DC}^{\rm one-loop}=
\frac{L^{6}T}
{48\pi^{3}r_{0}^{3}V}
\left(
-17
-7\gamma_{\mathrm{E}}
+22\ln 2
-15\ln 3
+7\ln\!\left(\frac{\pi l^{2}T}{r_{0}}\right)
\right),
\end{equation}
where we have dropped the $\ln \tilde{\epsilon}$ term, assuming a minimal-subtraction scheme.

First, $\sigma_{DC}$ goes to zero as the temperature approaches zero, $T\to 0$. This behavior is quite different from that obtained in the two-dimensional effective approach, where $\sigma_{DC}$ grows as $1/\sqrt{CT}$ at low temperatures. There is, however, some qualitative resemblance to the effective 2d result in the $CT\gg 1$ regime. The 2d result is dominated by terms that grow as ${\cal O}(T)$ and ${\cal O}(T^2)$, while the 4d result behaves as ${\cal O}(T)$ and ${\cal O}(T\log T)$, implying some qualitative agreement.

%%%%%%%%%%%%%%%%%%%%%%%%%%%%%%%%%%%%%%%%%%%%%%%%%%%%%%%%%%%%%%%
\subsection{Comparing the Schwarzian and the 4d gravitational path integral approaches}
\begin{figure}[htb]
    \centering
    \begin{subfigure}{\textwidth}
        \centering
        \includegraphics[width=0.75\linewidth]{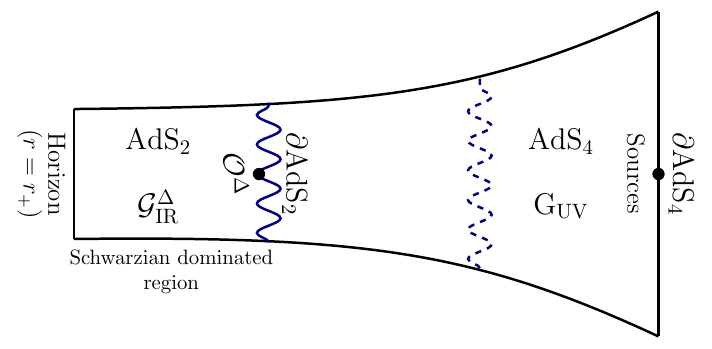}
        \caption{}
        \label{fig:4d-2d-comparison-2d}
    \end{subfigure}
    \begin{subfigure}{\textwidth}
        \centering
        \includegraphics[width=0.75\linewidth]{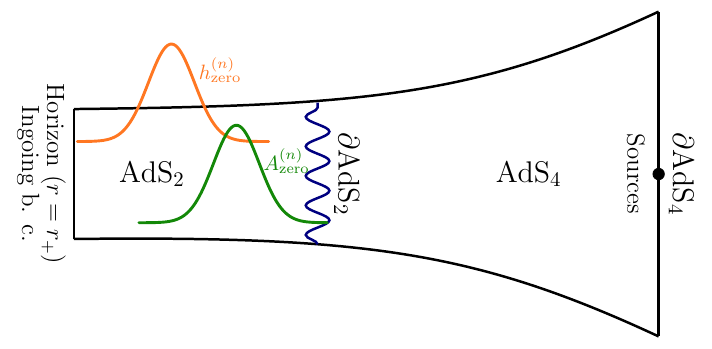}
        \caption{}
        \label{fig:4d-2d-comparison-4d}
    \end{subfigure}
    \caption{Schematic depiction of the 4d and 2d approaches to transport.}
    \label{fig:4d-2d-comparison}
\end{figure}

%%%%%%%%%%%%%%%%%%%%%%%%%%%%%%%%%%%%%%%%%%%%%%%%%%%%%%%%%%%%%%%
We have performed two different computations of the holographic conductivity, leading to different answers. Rather than trying to explain the difference,  here we will focus on clarifying the assumptions underlying each computation, as sketched in Figure~\ref{fig:4d-2d-comparison}. We leave a deeper analysis of the approximations to a separate publication.

%%%%%%

The 2d approach, schematically depicted in Fig.~\ref{fig:4d-2d-comparison-2d}, uses the nAdS$_2$/nCFT$_1$ correspondence to replace the throat region by the Schwarzian theory. In this approach, the fluctuation $\phi_{\rm bdy}$ at the boundary of AdS$_2$ is viewed as a source that couples the dual operator ${\cal O}_\Delta$ to the Schwarzian theory:
\begin{equation}
    S_{\rm Effective}^{1d}=S_{\rm Schwarzian}+\int dt \,\, \phi_{\rm bdy}(t)\,\,  {\cal O}_\Delta(t).
\end{equation}
It is plausible that this 2d approach is robust below the scale $T_q=1/C$. There are indications from exact results, such as the localization argument of \cite{Stanford:2017thb}, even though this argument itself does not extend to two-point functions.

For the 4d approach (see Fig.~\ref{fig:4d-2d-comparison-4d}), it is natural to expect that its regime of validity hovers around $T_q$ but does not extend below this value, because it is essentially a perturbative one-loop calculation, $T\gtrsim T_q$.

Let us further discuss other sources of discrepancies between the two approaches. In the 2d approach, the quantum fluctuations in the throat are described by the Schwarzian modes $\varepsilon_n$ (see, for example, \cite{Mertens:2022irh}). In the 4d approach, the fluctuations in the throat region are described by the zero modes $h_{\rm zero}^{(n)}$ and $A_{\rm zero}^{(n)}$. The precise couplings of these modes to the holographic fluctuations needed to compute the retarded Green's functions are certainly different. We surmise that a precise analysis of this point will reveal the origin of the differences between the two approaches. More precisely, in the 2d approach, the bulk electromagnetic fluctuation was reduced to a simple neutral scalar corresponding to $\Delta=2$. In the 4d approach, the situation was much more nuanced and involved a metric fluctuation in addition to the electromagnetic one. Some systematic observations regarding the 2d and 4d approaches at the classical level have been made in \cite{Castro:2021csm, Castro:2025pst}, a systematic quantum study along these lines remains open.

It is worth pointing out that the connection between the Schwarzian modes $\varepsilon_n$ and the higher-dimensional would-be zero modes has been established in a number of situations, including 4d Kerr black holes \cite{Kapec:2023ruw, Rakic:2023vhv} and a fairly universal class of backgrounds in various dimensions and with various asymptotics \cite{PandoZayas:2026vbg}. In all these cases, the final results from the 2d and higher-dimensional approaches are identical. The subtleties therefore arise when one considers the couplings of various holographic bulk modes to the would-be zero modes. We hope to discuss this problem systematically elsewhere.
%%%%
\iffalse
%%%%%%
    \begin{figure}[t]
        \centering
        \includegraphics[width=0.7\linewidth]{4d-2d-framework.pdf}
        \caption{Schematic depiction of the 4d and 2d approaches to transport.}
        \label{fig:optical_conductivity_frequency}
    \end{figure}
%%%%
\fi
%%%%%%%%%%%%%%%%%%%%%%%%%%%%%%%%%%%%%%%%%%%    
\section{Conclusions}\label{Sec:Conclusions}

In this manuscript, we have investigated the effects of quantum fluctuations in the near-horizon region of near-extremal black holes on holographic conductivity. We pursued two approaches to incorporate quantum corrections: the first uses the low-energy Schwarzian effective theory in lower dimensions while the other directly evaluates the four-dimensional gravitational path integral at one-loop level. We find that quantum corrections substantially modify the low-temperature behavior of the holographic conductivity in both approaches.

Our main result follows from incorporating the strong quantum fluctuations of the near-extremal throat, described by the Schwarzian theory, into the IR Green's function. The classical result in Equation \eqref{Eq:GIR} is thereby replaced by the quantum-corrected expression in Equation \eqref{Eq:GIR-Schw}, whose low-temperature behavior differs qualitatively from its classical counterpart. In particular, at sufficiently low temperatures we find 
\begin{equation}
\sigma_{DC}\sim \frac{1}{\sqrt{CT}}.
\end{equation}
The corresponding resistivity therefore vanishes as $T\to 0$. The emergence of this behavior suggests the possibility of a new low-temperature phase driven by quantum fluctuations in the near-extremal throat. Moreover, because the scaling originates from the universal structure of the Schwarzian correlator, we expect this behavior to persist across a broad class of holographic models with near-AdS${}_2$ IR geometries. It would be important to clarify the physical interpretation of this divergent conductivity, in particular by identifying more direct signatures of a possible instability or phase transition.

The direct four-dimensional one-loop computation also produces substantial quantum corrections to the conductivity, although its detailed temperature dependence does not precisely agree with that obtained from the dimensionally reduced Schwarzian description. This discrepancy raises an important question concerning the relation between dimensional reduction and quantization. While there are compelling reasons to formulate the low-energy dynamics in terms of the dimensionally reduced theory, dimensional reduction at the quantum level is known to involve important subtleties \cite{Duff:1980qv}. Related issues play a central role, for example, in precision computations of logarithmic corrections to black-hole entropy in AdS/CFT \cite{Liu:2017vbl}. A systematic understanding of the relation between the two approaches, and of the regimes in which they should be expected to agree, remains an important open problem. Here, we have restricted ourselves to identifying and exploring the possible quantum corrections governing low-temperature transport and leave a systematic comparison of the approaches to be discussed elsewhere.

Our results suggest several natural directions for further investigation. The first is to extend the analysis to other transport coefficients and to more realistic holographic backgrounds. In particular, incorporating momentum relaxation is essential for obtaining finite DC conductivities in holographic systems that would otherwise be translation invariant. We report results in this direction in a companion paper \cite{Suman-Mom-Rel}.

A second important direction is to develop the full hydrodynamic description of these quantum-corrected near-extremal systems. This requires extending the present analysis to fluctuations with nonzero spatial momentum\footnote{We thank Blaise Gout\'eraux and Cl\'ement Supiot for informing us of their unpublished results showing very interesting behavior in this regime.}, $\vec{k}\neq 0$. Progress in this direction was made in \cite{Nian:2025oei} within a particular approximation scheme. A more direct treatment of the momentum-dependent correlation functions and transport coefficients could help address several questions that remain open in attempts to formulate holographic hydrodynamics at parametrically low temperatures.

More broadly, it would be interesting to connect the transport computations via Kubo formulas to properties of low-temperature near-extremal black holes. For example, shear viscosity continues to be connected to the quantum-corrected absorption cross section of the black hole \cite{Emparan:2025sao,PandoZayas:2025snm,Cremonini:2025yqe,Kanargias:2025vul,Gouteraux:2025exs}. It would be interesting to determine whether such relations are manifestations of a more general connection between the quantum dynamics of black holes and low-temperature hydrodynamics. Establishing such a framework could provide the foundations for a {\it quantum membrane paradigm}, in which the transport properties of the dual system are directly encoded in the quantum dynamics of the near-horizon region.

 \acknowledgments{
 We are thankful to  Alejandra Castro,  Suman Das, Blaise Gout\'eraux, Xiao-Long Liu, Jun Nian, Cl\'ement Supiot, Joaqu\'in Turiaci, Misha Usatyuk, Cong-Yuan Yue, for insightful discussions. This work was partially supported by the U.S. Department of Energy under Grant No. DE-SC0007859 and by the National Natural Science Foundation of China (NSFC) under Grant No. 12247103. }

\appendix

\section{Wightman function evaluation}\label{App:Greens}

In this appendix, we provide some details of the approximation leading to the analytical approximation of the retarded Green's function from the Schwarzian theory.
	
\begin{figure}[t]
		\centering
		\begin{subfigure}{0.48\textwidth}
			\centering
			\includegraphics[width=\linewidth]{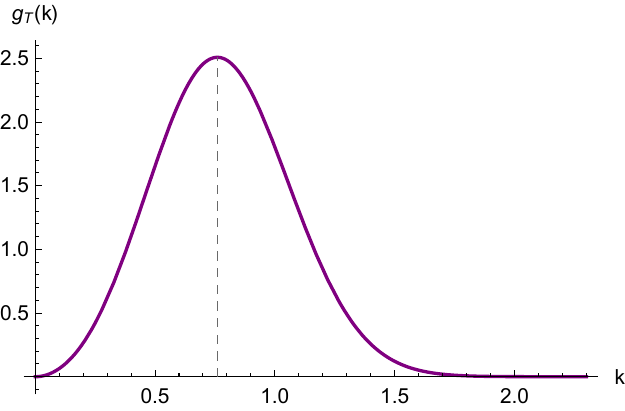}
			\caption{$C T = 0.1$}
		\end{subfigure}
		\begin{subfigure}{0.48\textwidth}
			\centering
			\includegraphics[width=\linewidth]{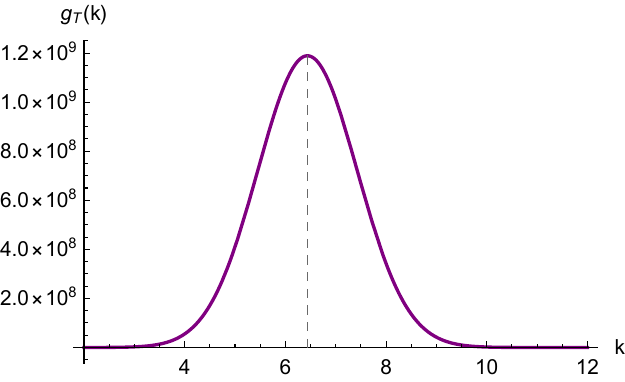}
			\caption{$C T = 1.0$}
		\end{subfigure}
		\begin{subfigure}{0.48\textwidth}
			\centering
			\includegraphics[width=\linewidth]{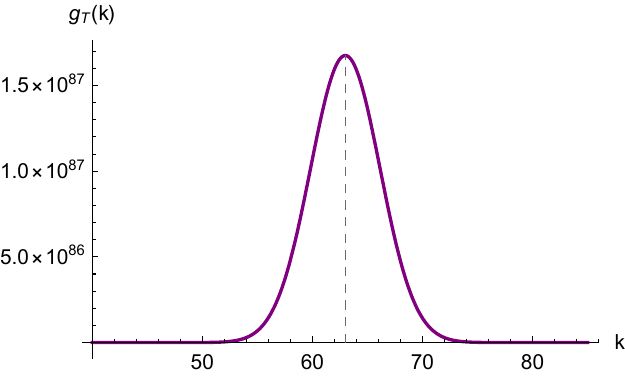}
			\caption{$C T = 10.0$}
		\end{subfigure}
		\caption{The function $g_{T}(k)$ is shown for different temperatures, $C T = 0.1, 1.0, 10.0$. In every case there is a bell-shaped region, but the maximum value increases rapidly with $T$.}
		\label{fig:g_and_f_image}
	\end{figure}

\begin{figure}[ht!]
		\centering
		\includegraphics[width=0.7\linewidth]{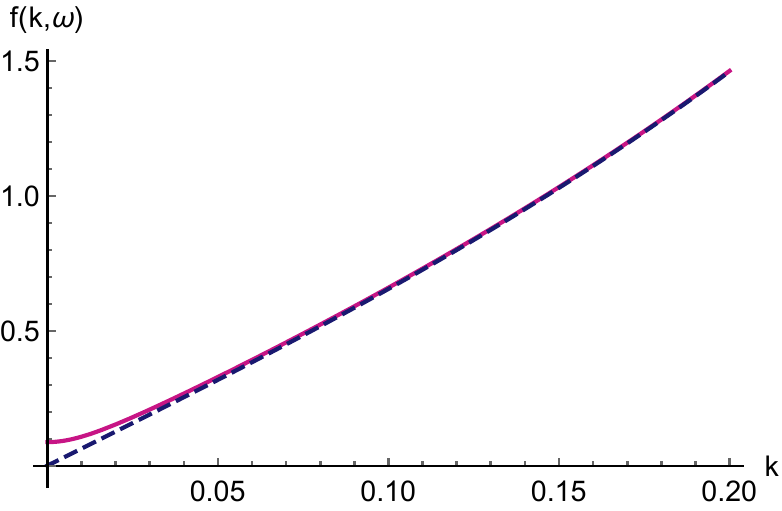}
		\caption{The function $f\left(k, \omega\right)$ (solid line) compared against its large $k$ approximation (dashed line), obtained for $C\omega = 10^{-4}$.}
		\label{fig:function_f}
	\end{figure}

\begin{figure}[t]
		\centering
		\begin{subfigure}{0.48\textwidth}
			\centering
			\includegraphics[width=\linewidth]{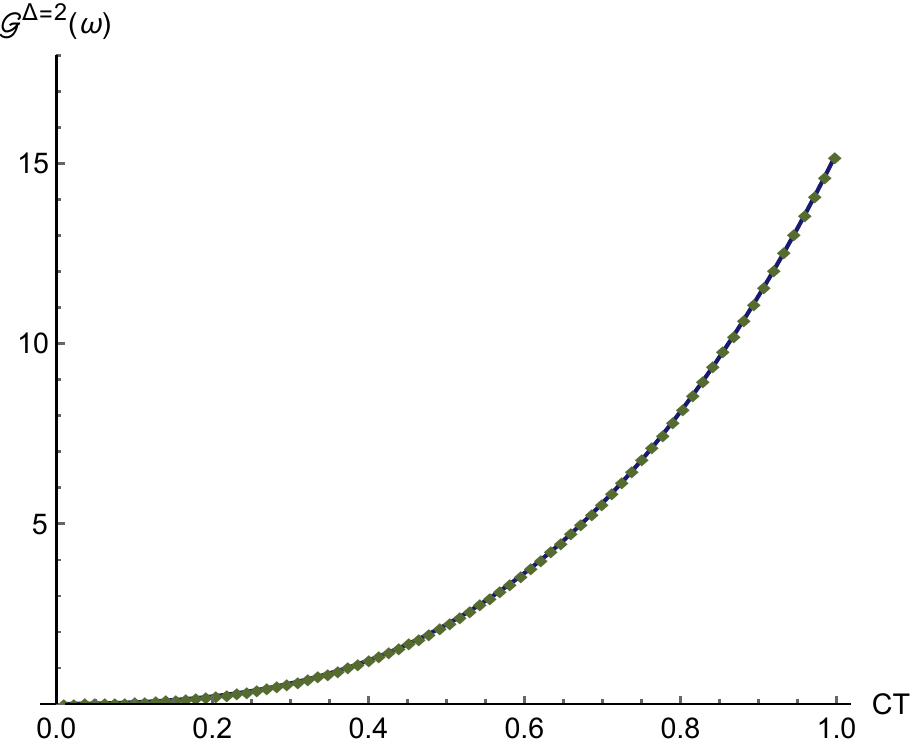}
			\caption{$C T \in \left[0.1, 1.0\right]$}
		\end{subfigure}
		\hfill
		\begin{subfigure}{0.48\textwidth}
			\centering
			\includegraphics[width=\linewidth]{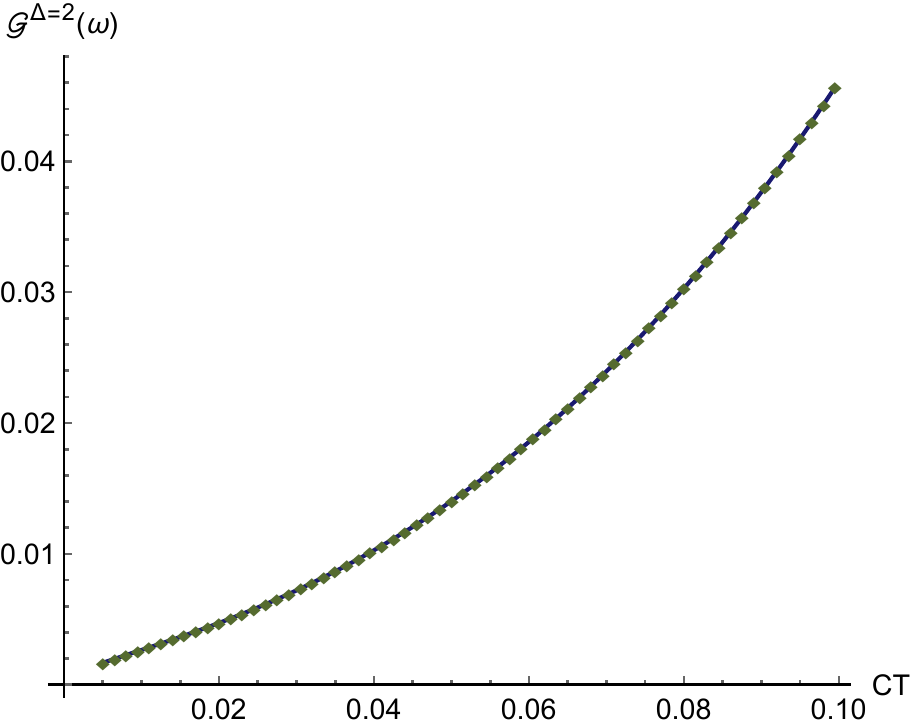}
			\caption{$C T \in \left[0.005, 0.1 \right]$}
		\end{subfigure}
		\caption{Comparison of the numerical results with analytic estimate of the two-dimensional Wightman function $\mathcal{G}^{\Delta=2}\left(\omega\right)$. The solid line denotes the analytical curve, and the dots denote numerically obtained values. For the numerical integration we choose $C\omega = 10^{-8}$.}
		\label{fig:WightmanG}.
	\end{figure}

	\noindent Let us rewrite the frequency space Wightman function as
	\begin{equation}
		\begin{split}
			\mathcal{G}^{\Delta} \left(\omega\right) &= \frac{e^{S_{0}} \left(2 C\right)}{Z(T)\, 4\pi^{4} \left(2 C\right)^{2\Delta} \Gamma\left(2\Delta\right)} \int_{0}^{\infty} dk\, g_{T}(k) \times f\left(k, \omega\right),\\
			\text{where}\quad g_{T}(k) &= k\, e^{-\frac{k^2}{2 C T}} \sinh \left(2\pi k\right),\\
			f\left(k, \omega\right) &= \prod_{\sigma_{1, 2} = \pm 1} \Gamma \left(\Delta + i\sigma_{1} \sqrt{k^2 + 2 C \omega} + i \sigma_{2} k \right) \sinh\left(2\pi \sqrt{k^{2} + 2 C \omega} \right).
		\end{split}
	\end{equation}
	The function $g_{T}(k)$ vanishes quadratically as $k \to 0$
	\begin{equation}
		g_{T} \left(k \right) = 2\pi k^{2} + O\left(k^{4} \right),
	\end{equation}
and it decays to zero exponentially as $k \to \infty$. This function has a bell-shaped region with a maximum at $k = k_{\ast}$, given by
	\begin{equation}
		\frac{k^{2}_{\ast}}{C T} - 1 = 2\pi k_{\ast} \coth \left(2\pi k_{\ast} \right).
	\end{equation}
The peak of $g_{T}\left(k\right)$ moves to higher $k$ as $T$ is increased, see Figure \ref{fig:g_and_f_image} for a visual representation. 
On the other hand, the function $f(k, \omega)$ is regular as $k \to 0$
	\begin{equation}
		\lim_{k \to 0} f\left(k, \omega \right) = 4 C \omega  (2 \pi  C \omega +\pi )^2 \coth \left(\sqrt{2 \pi^2  C \omega}\right),
	\end{equation}
and it diverges polynomially for large $k$
	\begin{equation} \label{eq:f_large_k}
		\lim_{k \to \infty} f\left(k, \omega\right) = 8\pi k^{3}\left(1 + \frac{\pi C \omega}{k} + \frac{2 C \omega  \left(2 \left(3+\pi ^2\right) C \omega + 9\right) + 3}{12 k^2} + \mathcal{O}\left(\frac{1}{k^3}\right)\right).
	\end{equation}
This function plays a central role in the evaluation of the integral, it is displayed in Figure \ref{fig:function_f}.
The function $g_{T}(k)$ always has a bell-shaped region, outside which it is highly suppressed. Following the approach in \cite{Kanargias:2025vul}, we observe that when $C\omega \ll CT$, the maximum of the bell-shaped function $g_{T}(k)$ lies within the polynomial regime of $f(k)$. This allows for the simplification of $f(k, \omega)$ via its large-$k$ asymptote \eqref{eq:f_large_k}. While this justification weakens for $CT \ll 1$, the authors of \cite{Kanargias:2025vul} demonstrated that the resulting error remains negligible. Consequently, we adopt this approximation throughout our analysis without further error quantification. We present a comparison of the analytic estimate for $\mathcal{G}^{\Delta = 2}\left(\omega \right)$ against the completely numerical result in Figure \ref{fig:WightmanG}.\\

\noindent After carrying out the integration using the large $k$ approximation, we obtain
{\scriptsize
\begin{equation} \label{eq:full_schwarzian_correlator}
    \begin{split}
       \mathcal{G}^{\Delta = 2} \left(\omega \right) = \frac{1}{144 C^{3}} &\Bigg(6 C \omega  (2 C \omega +1)^2 + 24 C^2 T \omega  \left(4 \pi ^2 C T+3\right) + \frac{24 \sqrt{2} e^{-2 \pi ^2 C T}}{\pi ^{3/2}} \left(4 \pi ^2 C T+5\right) (C T)^{3/2} \\ & + \frac{12 C T}{\pi^2} \left(8 \pi ^2 C T \left(2 \pi ^2 C T+3\right)+3\right) \mathrm{erf}\left(\sqrt{2} \pi  \sqrt{C T}\right) \\ & + \frac{e^{-2 \pi ^2 C T}}{\pi^2} \left(2 C \omega  \left(2 \left(3+\pi ^2\right) C \omega +9\right)+3\right) \left(e^{2 \pi ^2 C T} \left(4 \pi ^2 C T+1\right) \mathrm{erf}\left(\sqrt{2} \pi  \sqrt{C T}\right)+2 \sqrt{2 \pi } \sqrt{C T}\right) \Bigg).
    \end{split}
\end{equation}
}
We use this result in Section \ref{sec:quantum_from_2D} to write the low and high temperature expansions in Equation \eqref{eq:anest_high_low_T}.
%%%%%%
%%%%%

\begin{figure}[htb]
    \centering
    \includegraphics[width=0.6\linewidth]{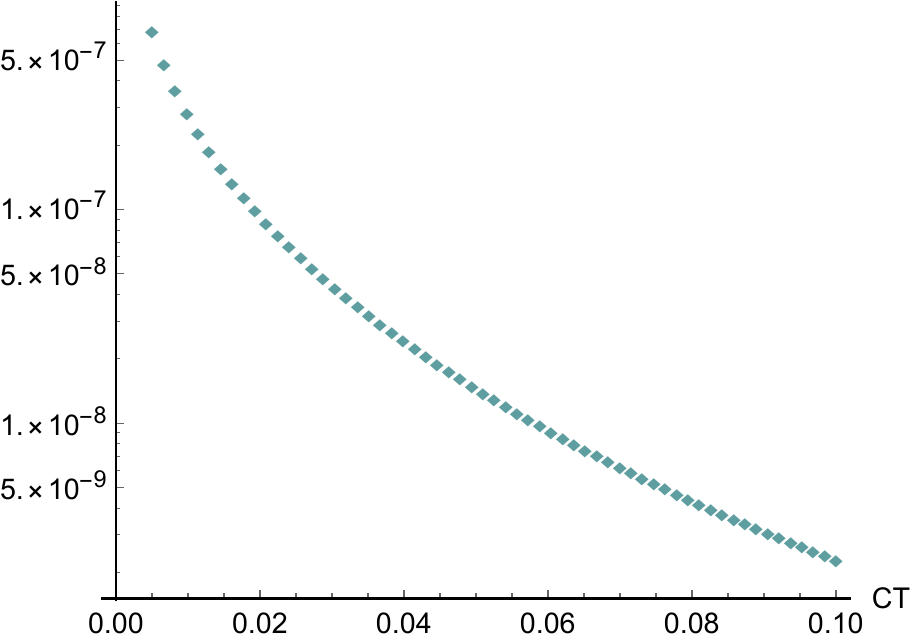}
    \caption{The relative error between numerical and analytical answers for $\mathcal{G}^{\Delta = 2}\left(\omega \right)$ for $C T \in \left[0.005, 0.1 \right]$. The $Y$-axis is drawn in $\log$ scale.}\label{Fig:Error-Relative}
\end{figure}

After comparison between the numerical and analytical results, we observe that the relative error $\left|\frac{\mathcal{G}^{\Delta=2}_{\mathrm{num}} - \mathcal{G}^{\Delta=2}_{\mathrm{est}}}{\mathcal{G}^{\Delta=2}_{\mathrm{num}}} \right|$ is $O\left(10^{-7} \right)$ for the smallest temperature value considered in this computation, and the error further decreases as the temperature is increased.  The precise plot of the relative error is presented in Figure \ref{Fig:Error-Relative}.

%%%%%%
%%%%%%

Therefore, despite the reduced validity of our approximation for small $CT$, we may trust the analytic estimate with tolerable accuracy.

%%%%%%%%%%%%%%%%%%%%%%%%%%%%%%%%%%%%%%%%%%%%%%%%%%
\section{Details of the zeta function regularization}
\label{App: 4D details}
In this appendix, we provide details on the Zeta function regularization used in calculating the one-loop part of the partition function.

To obtain a finite answer for the infinite product \eqref{eq:infinite product}, we are going to use Zeta function to regularize the infinite product of the form
\begin{equation}
Z(\epsilon_1,\epsilon_2)
:=
\prod_{n=n_0}^{\infty}
\left(dn+\epsilon_1\epsilon_2 q_n\right)^{-1},
\qquad
n_0=1,2,
\end{equation}
where
\begin{equation}
q_n
=
an^2+bn+c+\alpha_1 n\psi(n)+\alpha_2 n^2\psi(n),
\end{equation}
In calculating the infinite product of eigenvalues, we have $\epsilon_1\equiv\epsilon C_{\rm in}(\omega),\,\epsilon_2\equiv\epsilon C_{\rm in}(-\omega)$.
The $dn$ term comes from $\Lambda(g^{(1)},A^{(1)})$ and the $q_n$ term corresponds to $\Lambda(\delta g,\delta A)$. The $n^2$ terms are not necessary in the calculation of the imaginary part of the Green's function. We include them because such terms appear in the real part of the lifted eigenvalues. The Zeta function regularization formula derived in this appendix will also apply if one also wants to include the real part of the lifted eigenvalues.
Define $t\equiv\epsilon_1\epsilon_2.$
For \(d>0\) and \(\mathrm{Re}\,s\) sufficiently large,
\begin{equation}
(dn+tq_n)^{-s}
=
(dn)^{-s}
-
s t d^{-s-1}q_n n^{-s-1}
+
O(t^2).
\end{equation}
Thus
\begin{equation}
\sum_{n=n_0}^{\infty}(dn+tq_n)^{-s}
=
d^{-s}\sum_{n=n_0}^{\infty}n^{-s}
-
s t d^{-s-1}B_{n_0}(s)
+
O(t^2),
\end{equation}
where
\begin{equation}
B_{n_0}(s)
:=
\sum_{n=n_0}^{\infty}q_n n^{-s-1}.
\end{equation}

We use the meromorphic continuations
\begin{equation}
P_1(s):=\sum_{n=1}^{\infty}\psi(n)n^{-s}
=
-\frac{1}{2s}
+\frac12
+
O(s),
\end{equation}
and
\begin{equation}
P_2(s):=\sum_{n=1}^{\infty}\psi(n)n^{1-s}
=
-\frac{1}{12s}
+\frac38
+
O(s).
\end{equation}
Also,
\[
\zeta(s+1)=\frac1s+\gamma_{\mathrm{E}}+O(s),
\qquad
\zeta(-1)=-\frac{1}{12},
\qquad
\zeta(0)=-\frac12.
\]

\subsection*{Case \(n_0=1\)}

For \(n_0=1\),
\begin{equation}
B_1(s)
=
a\zeta(s-1)
+
b\zeta(s)
+
c\zeta(s+1)
+
\alpha_1 P_1(s)
+
\alpha_2 P_2(s).
\end{equation}
Therefore
\begin{equation}
B_1(s)
=
\frac{1}{s}
\left(
c-\frac{\alpha_1}{2}-\frac{\alpha_2}{12}
\right)
-
\frac{a}{12}
-
\frac{b}{2}
+
c\gamma_{\mathrm{E}}
+
\frac{\alpha_1}{2}
+
\frac{3\alpha_2}{8}
+
O(s).
\end{equation}
Since
\[
d^{-s-1}
=
\frac1d
\left(1-s\log d+O(s^2)\right),
\]
we obtain
\begin{equation}
\log Z_1(t)
=
\log Z_1(0)
+
\frac{t}{d}K_1
+
O(t^2),
\end{equation}
where
\begin{equation}
K_1
=
\frac{a}{12}
+
\frac{b}{2}
+
c(\log d-\gamma_{\mathrm{E}})
-
\frac{\alpha_1}{2}(1+\log d)
-
\alpha_2
\left(
\frac38+\frac{\log d}{12}
\right).
\end{equation}
The zeroth-order factor is
\begin{equation}
Z_1(0)
=
\prod_{n=1}^{\infty}(dn)^{-1}
=
\sqrt{\frac{d}{2\pi}}.
\end{equation}
Hence
\begin{equation}
Z_1(\epsilon_1,\epsilon_2)
=
\sqrt{\frac{d}{2\pi}}
\left[
1+
\frac{\epsilon_1\epsilon_2}{d}
K_1
+
O\!\left((\epsilon_1\epsilon_2)^2\right)
\right].
\label{eq:Z1-digamma-result}
\end{equation}

\subsection*{Case \(n_0=2\)}

For \(n_0=2\), we subtract the \(n=1\) term. Since
\[
\psi(1)=-\gamma_{\mathrm{E}},
\]
we have
\[
q_1=a+b+c-\gamma_{\mathrm{E}}(\alpha_1+\alpha_2).
\]
Equivalently,
\begin{equation}
B_2(s)
=
B_1(s)-q_1.
\end{equation}
Thus
\begin{equation}
B_2(s)
=
\frac{1}{s}
\left(
c-\frac{\alpha_1}{2}-\frac{\alpha_2}{12}
\right)
-
\frac{13a}{12}
-
\frac{3b}{2}
+
c(\gamma_{\mathrm{E}}-1)
+
\alpha_1
\left(
\frac12+\gamma_{\mathrm{E}}
\right)
+
\alpha_2
\left(
\frac38+\gamma_{\mathrm{E}}
\right)
+
O(s).
\end{equation}
Therefore
\begin{equation}
\log Z_2(t)
=
\log Z_2(0)
+
\frac{t}{d}K_2
+
O(t^2),
\end{equation}
where
\begin{equation}
K_2
=
\frac{13a}{12}
+
\frac{3b}{2}
+
c(1-\gamma_{\mathrm{E}}+\log d)
-
\alpha_1
\left(
\gamma_{\mathrm{E}}+\frac{1+\log d}{2}
\right)
-
\alpha_2
\left(
\gamma_{\mathrm{E}}+\frac38+\frac{\log d}{12}
\right).
\end{equation}
The zeroth-order factor is
\begin{equation}
Z_2(0)
=
\prod_{n=2}^{\infty}(dn)^{-1}
=
\frac{d^{3/2}}{\sqrt{2\pi}}.
\end{equation}
Hence
\begin{equation}
Z_2(\epsilon_1,\epsilon_2)
=
\frac{d^{3/2}}{\sqrt{2\pi}}
\left[
1+
\frac{\epsilon_1\epsilon_2}{d}
K_2
+
O\!\left((\epsilon_1\epsilon_2)^2\right)
\right].
\label{eq:Z2-digamma-result}
\end{equation}

The two coefficients are related by
\begin{equation}
K_2=K_1+q_1,
\qquad
q_1=a+b+c-\gamma_{\mathrm{E}}(\alpha_1+\alpha_2).
\end{equation}

By substituting corresponding coefficients $\{a,b,c,\alpha_1,\alpha_2\}$, one can calculate the one-loop part of the partition function. Using this regularization procedure, one can also see that the real and imaginary parts do not mix after the infinite products. This justifies keeping only the imaginary part of the lifted eigenvalues. 

\bibliographystyle{JHEP}
\bibliography{references}

\end{document}